\documentclass[12pt,a4paper]{article}
\usepackage{jheppub}
\usepackage{amssymb}
\usepackage{amsmath}
\usepackage{mathtools}
\usepackage{graphicx}
\usepackage{enumitem}
\usepackage{xcolor}
\usepackage[utf8]{inputenc}
\usepackage{slashed}
\usepackage{amsmath}
\usepackage{bm}
\usepackage{caption}	
\usepackage{enumerate}
\usepackage{braket}
\usepackage{comment}

\usepackage{chngcntr}

\makeatletter
\newcommand{\raisemath}[1]{\mathpalette{\raisem@th{#1}}}
\newcommand{\raisem@th}[3]{\raisebox{#1}{$#2#3$}}
\makeatother

\def\be{\begin{equation}}
\def\ee{\end{equation}}
\def\bea{\begin{eqnarray}}
\def\eea{\end{eqnarray}}
\def\beal{\begin{equation}\begin{aligned}}
\def\eeal{\end{aligned}\end{equation}}

\def\cM{\mathcal{M}}
\def\cA{\mathcal{A}}

\def\angle#1{\langle #1 \rangle}
\def\Angle#1{\left\langle #1 \right\rangle}

\def\be{\begin{equation}}
\def\ee{\end{equation}}
\def\bea{\begin{eqnarray}}
\def\eea{\end{eqnarray}}

\def\bep{\boldsymbol{\mathbf{\ep}}}
\def\ep{\varepsilon}

\def\OliSpin{\cite{Schlotterer:2010kk}}

\def\OurKerr{\cite{Chiodaroli:2021eug}}
\def\mshA{\angle{\bm1\bm2}}
\def\mshS{[\bm1\bm2]}

\def\pH{k}
\def\epH{{\ep_{k}}}

\def\kdota#1{\pH \cdot a_{(#1)}}
\def\kdota#1#2{(\pH \cdot a_{(#1)})^{#2}}

\title{Classical Limit of Higher-Spin String Amplitudes}
\author[a,b]{Lucile Cangemi,} 
\author[a,b]{Paolo Pichini\,}
\affiliation[a]{Uppsala Universitet, \\ 
Department of Physics and Astronomy, Uppsala University,\\
Box 516, 75120 Uppsala, Sweden}
\affiliation[b]{Kavli Institute for Theoretical Physics,\\
University of California, Santa Barbara, CA, 93106-4030, USA.
}
\emailAdd{lucile.cangemi@physics.uu.se}
\emailAdd{paolo.pichini@physics.uu.se}

\abstract{It has been shown that a special set of three-point amplitudes between two massive spinning states and a graviton reproduces the linearised stress-energy tensor for a Kerr black hole in the classical limit. In this work we revisit this result and compare it to the analysis of the amplitudes describing the interaction of leading Regge states of the open and closed superstring. We find an all-spin result for the classical limit of two massive spinning states interacting with a photon or graviton. This result differs from Kerr and instead matches the current four-vector and the stress-energy tensor generated by a classical string coupled to electromagnetism and gravity respectively. For the superstring amplitudes, contrary to the black-hole case, we find that the spin to infinity limit is necessary to reproduce the classical spin multipoles.}

\preprint{UUITP-30/22}

\begin{document}
\maketitle

\section{Introduction}

Massive higher-spin scattering amplitudes have received a lot of attention in the past decade for their role in calculating classical gravity observables. Early work by ref.~\cite{Vaidya:2014kza} suggested that amplitudes between low-spin massive states and gravitons could be used to compute the lowest orders in the spin-multipole expansion of black hole observables, such as scattering angles and gravitational potentials. Later an all-order in spin result was obtained when a special set of arbitrary-spin three-point amplitudes in ref.~\cite{Arkani-Hamed:2017jhn} was used to calculate the linearised energy-momentum tensor of a Kerr black hole, and therefore extract the conservative dynamics of a binary system of spinning black holes at leading order in the post-Minkowskian (PM) expansion~\cite{Vines:2017hyw,Guevara:2017csg,Guevara:2018wpp,Chung:2018kqs,Guevara:2019fsj,Chung:2020rrz}.

The double copy,~\cite{Kawai:1985xq,Bern:2008qj,Bern:2010ue,Bern:2019prr}, and its extension to massive states, has proved to be a powerful tool when calculating the relevant gravitational amplitudes from their simpler Yang-Mills counterparts \cite{Johansson:2015oia,Ochirov:2018uyq,Johansson:2019dnu,Chung:2019duq, Bautista:2019evw, Arkani-Hamed:2019ymq,Edison:2020ehu,Johnson:2020pny,Momeni:2020hmc,Momeni:2020vvr,Haddad:2020tvs,Emond:2020lwi,Bjerrum-Bohr:2020syg,Brandhuber:2021kpo}. This field theoretic relation leaves an imprint in the classical theory in the form of the classical double copy, which was first introduced by ref.~\cite{Monteiro:2014cda} to relate the structure of Schwarzschild and Kerr black hole solutions to corresponding solutions in classical electrodynamics. Conversely, tools used to relate classical solutions, such as the Newman-Janis shift, have since been exported to the amplitudes framework providing simplifications in the treatment of spin~\cite{Huang:2019cja,Chung:2019yfs,Guevara:2020xjx}.

The classically relevant spin information can be extracted after the introduction of the Pauli-Lubanski spin operator, which connects quantum and classical spin~\cite{Maybee:2019jus}. Amplitudes can be expressed as polynomials in the expectation value of this operator, known as the \textit{spin multipole} expansion~\cite{Vaidya:2014kza}, a process that has been facilitated by the use of the on-shell spinor-helicity variables introduced by ref.~\cite{Arkani-Hamed:2017jhn} and their extension, the on-shell HPET variables introduced in ref.~\cite{Aoude:2020onz}. The general approach to the classical limit of amplitudes can be summarised as taking the simultaneous $\hbar \to 0$ and quantum spin $s \to \infty$ limit while fixing $\hbar s \sim \mathcal{O}(1)$ as introduced in ref.~\cite{Arkani-Hamed:2019ymq}. Recently work in ref.~\cite{Aoude:2021oqj} provided an alternative formulation in terms of spin-coherent states.
The three point amplitudes in ref.~\cite{Arkani-Hamed:2017jhn} exhibit special behaviour when expressed in terms of the spin variables described above. Any finite spin-$s$ amplitude reproduces the same spin multipole expansion that arises in the $s \to \infty$ limit, albeit only up to order $2s$ in the Pauli-Lubanski operator. This is referred to as \textit{spin universality}~\cite{Holstein:2008sw,Holstein:2008sx} and it makes it possible to extract classical observables from finite-spin amplitudes, up to a finite order in the spin multipole expansion. Moreover, the three-point Kerr amplitudes seem to have additional special features in the eikonal approximation including suppressed entanglement~\cite{Aoude:2020mlg,Chen:2021huj, Haddad:2021znf}.

Conservative observables for spinning black holes have since been studied in a variety of different frameworks, including higher-spin effective field theories and worldline quantum field theory~\cite{Bern:2020buy, Liu_2021, Jakobsen:2021lvp, Jakobsen:2022fcj, Riva:2022fru}. The current state of the art for spinning observables at NLO (one loop) corresponds to quartic order in spin~\cite{Guevara:2018wpp,Chung:2018kqs,Siemonsen:2019dsu,Damgaard:2019lfh,Bern:2020buy,Kosmopoulos:2021zoq,Chen:2021qkk, Menezes:2022tcs} and, at NNLO (two loop), quadratic order in spin~\cite{Jakobsen:2022fcj,FebresCordero:2022jts}.\footnote{The NLO quartic-in-spin results \cite{Guevara:2018wpp,Chen:2021qkk,Menezes:2022tcs} have not yet been conclusively shown to correspond to classical black holes, but they are consistent with available partial constraints from matching to self-force calculations \cite{Siemonsen:2019dsu}.} Moreover, recent work extends beyond black holes by including tidal deformations~\cite{Aoude:2020ygw, Bern:2020uwk} and going beyond conservative dynamics to the study of radiative observables, such as the waveform and the power emitted by a binary system~\cite{Mogull:2020sak,AccettulliHuber:2020dal,Cristofoli:2021vyo,Herrmann:2021lqe,Jakobsen:2021lvp,Mougiakakos:2021ckm,Bautista:2021inx,Alessio:2022kwv}.

Despite this progress, a complete understanding of the effective theory that can match Kerr observables to all orders is still missing \cite{Bekaert:2022poo}. In particular, the correct form of the gravitational Compton amplitude involving two massive spinning states and two gravitons, needed to extend the one-loop calculation to higher orders in spin, is still an open problem. A candidate Compton amplitude was computed in refs.~\cite{Arkani-Hamed:2017jhn,Johansson:2019dnu} using factorisation and BCFW recursion relations~\cite{Britto:2004ap,Britto:2005fq}, but the results developed spurious poles for any spin higher than two. Complementary approaches have emerged to tackle this problem. One seeks to resolve the ambiguity in the classical regime, extrapolating patterns that emerged in the one loop conservative observables for low spin to higher spins~\cite{Aoude:2022trd,Bern:2022kto,Aoude:2022thd}. For certain observables it was also shown that the details of the Compton amplitude can be ignored in the relevant regimes~\cite{Alessio:2022kwv, Chen:2022yxw}. Alternatively, recent work by one of the authors,~\cite{Chiodaroli:2021eug}, sought to exploit the many constraints and consistency conditions that are present in higher-spin theory and derived the Kerr three-point amplitudes from first principles, up to spin-$5/2$ massive states, from high-energy unitarity considerations. The proposed spin-$5/2$ Compton amplitude, also appearing in ref.~\cite{Falkowski:2020aso} from a novel massive BCFW-type shift, has several nice properties as demonstrated in ref.~\cite{Aoude:2022thd}. However, it seems to be at odds with spin universality at quartic order in spin. At this point in time a first principles calculation of the Kerr Compton amplitude is still missing, however some progress has been made by mapping this problem to plane-wave scattering in a black-hole background as described by the Teukolsky equation~\cite{Bautista:2021wfy}.

This work also starts from the observation in ref.~\cite{Chiodaroli:2021eug} that the three-point Kerr amplitudes \cite{Arkani-Hamed:2017jhn} are connected to highly constrained theories in the higher-spin literature. When looking for a well-behaved model of fundamental higher-spin particles that satisfies high-energy unitarity constraints, string theory is a natural place to consider. Various aspects of massive states in string theory have been studied for several decades, including work on properties of the spectrum and vertex operators~\cite{KOH1987201,Berkovits:2002qx,Hanany:2010da,Bianchi:2010es,Lust:2012zv,Lee:2012fu,Feng:2012bb,Fu:2013xba}, and amplitude computations that probe the interaction of massive and massless string states~\cite{PhysRevD.83.046005,Feng:2010yx,Schlotterer:2011zz,Feng:2011qc,Boels:2012if,Tsulaia:2012rb,Feng:2012qia,Boels:2014dka,Minahan:2014usa,Hansen:2015pqa,Chakrabarti:2018bah,Lee:2019ufv,Gross:2021gsj,Jusinskas:2021bdj,Guillen:2021mwp}. In particular, the interaction of massive string states with gravity was the subject of several studies~\cite{Giannakis:1998wi,Buchbinder:1999be,Buchbinder:1999ar,Buchbinder:2000ta}. Massive higher-spin particles in flat space are subject to no-go theorems and high-energy pathologies (see for example refs. \cite{PhysRev.186.1337,PhysRev.188.2218}), hence studying massive string states offered valuable insight on how such issues can be circumvented~\cite{Taronna:2010qq,Joung:2012rv,Taronna:2012gb,Rahman:2015pzl,Marotta:2021oiw}. Exploring the high energy regime, where the higher-spin tower becomes most relevant, also leads to interesting observations that hint to a spontaneous-symmetry-breaking-like mechanism responsible for the particles' mass~\cite{Gross:1988ue,Lee:1994fy,Chan:2003ee,Chan:2004tb,Chan:2005ne,Chan:2005zp,Chan:2006qf,Polyakov:2009pk,Polyakov:2010qs,Fotopoulos:2010ay,Bianchi:2011se,Lee:2019lvp,Gross:2021gsj}. An interesting example can be found in ref.~\cite{Sagnotti:2010at}, where the three-point amplitudes for massive states of the bosonic string are computed and uplifted to off-shell vertices to highlight their high-energy properties. Similarly ref.~\cite{Schlotterer:2010kk} provided general-spin amplitudes for leading Regge states of the open and closed superstring.

In this work, we study the leading Regge superstring amplitudes, in particular the three-point amplitudes between two massive spinning states and a photon/gluon or graviton, and apply the same classical limit technology that was introduced for the Kerr case. The resulting amplitudes, expanded in spin multipoles, are shown to agree with Kerr at the first mass level (quantum spin $s=4$, in the closed string case). However, for higher spin numbers, the results deviate from Kerr and, in the infinite-spin limit, they match the electromagnetic current and stress-energy tensor of known classical string solutions \cite{Ademollo:1974te,Lust1989,zwiebach_2004}, manifesting a novel classical double copy relation between the two. The purpose of this work is twofold. First and foremost, it provides an application of the classical limit to a different set of amplitudes, highlighting important elements that are easily overlooked when studying the Kerr case. For instance, whereas the three-point Kerr amplitudes obey spin universality, the string amplitudes do not, such that the spin multipoles associated to the classical solutions can be reproduced only after taking the infinite-spin limit. Furthermore, this work paves the way to future investigations of the classical limit of string amplitudes. This includes exploring other string theories, such as bosonic and heterotic strings, as well as going beyond the leading Regge trajectory, including treatment of subleading trajectories as well as string coherent states~\cite{Bianchi:2019ywd}, in the hope that there exists a set of string amplitudes that generate the correct classical amplitudes for Kerr black holes.

The structure of this paper is as follows. In section \ref{sect:recap} we define spin vector variables and review their application to classical limits of amplitudes, with emphasis on the Kerr black-hole case. In section \ref{sect:stringampl} we apply the same technology to the string amplitudes between two leading Regge trajectory states and a photon or graviton, highlighting some conceptual differences to the black-hole case and computing the all-spin classical limit. This is shown to match some standard classical string solutions in section \ref{sect:stringclass}. We conclude in section \ref{sect:conclusion} discussing our results and outlining some interesting future directions. In the appendix, we provide more detail on the classical string solutions we studied.


\section{Spin Variables and Classical Limits}
\label{sect:recap}
In this section we will review the formalism for the classical limit of three-point spinning amplitudes in the spirit of ref. \cite{Arkani-Hamed:2019ymq}, presenting it in a form that is best suited to this work. We will start by introducing the spin variables that make the spin dependence in the amplitudes explicit.

\subsection{Spin vector variables}

Let us consider a spin-$s$ free particle in a quantum field theory, with momentum $p_1$ and mass $m$, satisfying the on-shell condition $p_1^2 = m^2$. This can arise as an asymptotic state in a scattering amplitude and, as we will explain in more detail, it will be identified with a black hole or another classical object upon taking the classical limit. Its intrinsic spin is described by the expectation value of the Pauli-Lubanski operator\footnote{We use the mostly-minus metric and the following conventions for the Levi-Civita tensor $\epsilon^{0123} =1$ and the Pauli matrices, $\sigma^\mu = (1, \sigma^1, \sigma^2, \sigma^3)$, $\bar{\sigma}^\mu = (1, -\sigma^1, -\sigma^2, -\sigma^3)$. We also define $\sigma_{\mu \nu}$ either as ${\left(\sigma_{\mu \nu} \right)_{\alpha}}^{\beta}={\left(\sigma_{[\mu} \bar{\sigma}_{\nu]}\right)_{\alpha}}^{\beta}$ or  ${\left(\sigma_{\mu \nu} \right)^{\dot \alpha}}_{\dot \beta}={\left(\bar{\sigma}_{[\mu} \sigma_{\nu]}\right)^{\dot \alpha}}_{\dot \beta}$ acting on the Weyl spinors $|1\rangle_\beta$ or $|1]^{ \dot \beta}$ respectively, where the antisymmetrisation convention includes a factor of $1/2!$.}
\begin{equation} \label{eq:spinopdef}
    S^\mu_{(s)} = \frac{1}{2m} \epsilon^{\mu \nu \rho \sigma} p_{1 \nu} M_{(s) \rho \sigma},
\end{equation} 
where the subscript $(s)$ indicates inserting the spin-$s$ representation of the Lorentz generators,
\begin{equation}
    {(M_{(s) \mu \nu})^{\alpha(s)}}_{\beta(s)} = \begin{cases} 2 i s \, {\delta^{(\alpha_1}}_{[\mu} \eta_{\nu] (\beta_1} \delta^{\alpha_2}_{\beta_2} \dots \delta^{\alpha_s)}_{\beta_s)} &\textrm{ for integer spin,}\vspace{3mm} \\ 
     i s \, {(\sigma_{\mu\nu})_{( \alpha_1}}^{( \beta_1} \delta^{ \alpha_2}_{ \beta_2} \dots \delta^{ \alpha_{2s})}_{ \beta_{2s})}&\textrm{ for half-integer spin}.
    \end{cases}
\end{equation} 
The multi-indices $\alpha(s), \beta(s)$ represent the fully-symmetrised $s$ vector ($2s$ spinor) indices in the (half-) integer spin case, where we include a factor of $1/s!$ ($1/2s!$) in the symmetrisation. We use massive Weyl spinors to describe half-integer spins and we provide formulae for the right-handed spinors, but equivalent results for left-handed spinors can be obtained by conjugation. The Weyl spinors can also be used for integer spins, but in that case we choose to work with covariant polarisation vectors.

In section \ref{sec:Kerr} we will relate the operator in eq. \eqref{eq:spinopdef} to a classical notion of spin. In order to do so, we must first introduce the \textit{symmetrised} expectation value of $S_{(s)}^\mu$,
\begin{equation} \label{eq:symmexpval}
    \angle{S_{(s)}^{(\mu_1} \dots S_{(s)}^{\mu_n)}} \\
    =\begin{cases} \frac{1}{(\bar{\bep}_1 \cdot \bep_1)^{s}}\bar{\bep}_{1 \alpha(s)} {(\Sigma^{\mu_1 \dots \mu_n}_{(s)})^{\alpha(s)}}_{\beta(s)} \bep_1^{\beta(s)}, &s\in \mathbb{N} \\
    \frac{1}{\angle{\bar{\mathbf{1}} \mathbf{1}}^{2s}}\langle \bar{\mathbf{1}}|^{\alpha(s)}  {(\Sigma^{\mu_1 \dots \mu_n}_{(s)})^{\alpha(s)}}_{\beta(s)} |\mathbf{1}\rangle_{\beta(s)},  &s\in \frac{1}{2}\mathbb{N}
    \end{cases}
\end{equation} 
where $ {\Sigma_{(s)}^{\mu_1 \dots \mu_n \alpha(s)}}_{\raisemath{3pt}{\beta(s)}} = 
{(S_{(s)}^{(\mu_1} S_{(s)}^{\mu_2}\dots S_{(s)}^{\mu_n)})^{\alpha(s)}}_{\raisemath{3pt}{\beta(s)}} $ such that matrix multiplication between the spin operators is left implicit. This definition is quantum mechanical in spirit, where we take the expectation value with respect to a particle's associated in-state, $\bep_1$ or $|\mathbf{1} \rangle$, and out-state $\bar{\bep}_1$ or $|\bar{\mathbf{1}}\rangle$, which are related by complex conjugation. Here $\bep_1^{\alpha(s)} \equiv \bep_1^{\alpha_1} \dots \bep_1^{\alpha_s}$ is the polarisation tensor of the spin-$s$ particle considered. It is symmetric, traceless and transverse due to the properties $\bep_1^2 = \bep_1 \cdot p_1 = 0$, discussed below. Alternatively, the spin-$s$ particle can be described by the tensor product of $2s$ massive Weyl spinors, $|\bm1 \rangle_{\alpha(s)} \equiv |\bm1\rangle_{ \alpha_1} \dots |\bm1\rangle_{\alpha_{2s}}$.

These bolded massive spinors are related to the conventional massive spinors via
\begin{equation}
\label{eq:massivespinors}
    |\mathbf{1} \rangle_{ \alpha_1} = |1^a\rangle_{ \alpha_1} z_{1 a} , \hspace{5mm} |\bar{\mathbf{1}}\rangle_{\alpha_1} = |1^a\rangle_{\alpha_1} \bar{z}_{1 a}.
\end{equation} The $SU(2)$ index, $a$, corresponds to the little group of $p_1$. We contract the spinors with auxilary $SU(2)$ spinors $z_{1 a}$ and $\bar{z}_{1 a}$, which pick out a specific polarisation direction \cite{Chiodaroli:2021eug}. Such auxilary variables have a nice interpretation in the context of spin-coherent states \cite{Aoude:2021oqj}. 

The polarisation vectors can also be written in terms of the massive spinors,
\begin{equation} \label{eq:poldefn}
    \bep_1^\mu = \frac{\langle \mathbf{1} | \sigma^\mu |\mathbf{1}]}{\sqrt{2}m}  , \hspace{5mm} \bar{\bep}_1^\mu = -\frac{\langle \bar{\mathbf{1}}| \sigma^\mu |\bar{\mathbf{1}}]}{\sqrt{2}m} . 
\end{equation}
The definition \eqref{eq:massivespinors} implies that $\angle{\mathbf{1}\mathbf{1}}=\angle{\bar{\mathbf{1}}\bar{\mathbf{1}}}=0$, $\angle{\bar{\mathbf{1}} \mathbf{1}} = -m \epsilon^{a b} \bar{z}_{1 a} z_{1 b} $, in terms of the two-dimensional Levi-Civita tensor normalised by $\epsilon^{1 2} = 1$. As a consequence, $\bep_1^2 = \bar{\bep}_1^2 = \bep_1 \cdot p_1 = \bar{\bep}_1 \cdot p_1 = 0$ as claimed, and $\bar{\bep}_1 \cdot \bep_1 = - (\epsilon^{a b} z_{1 a} \bar{z}_{1 b} )^2 $.  For simplicity, we choose to work with the normalisation $\epsilon^{a b} \bar{z}_{1 a} z_{1 b} =-1$ such that $\bar{\bep}_1 \cdot \bep_1 = - 1$.   

Given the definitions above, we can write the spin vector expectation value in terms of these on-shell variables
\begin{equation} \label{eq:spindefn}
    \angle{S_{(s)}^\mu} 
    =- \frac{s}{m} \langle 1^a|\sigma^\mu |1^b] \bar{z}_{1 (a} z_{1 b)}
    = -\frac{i s}{m} \epsilon^{\mu \nu \rho \sigma} p_{1\nu} \bar{\bep}_{1\rho} \bep_{1\sigma} .
\end{equation} 

We will see in section \ref{sec:Kerr} that, in the $s \to \infty$ limit, these spin variables reduce to classical spin vectors and allow us to extract classical observables. 

However, rewriting scattering amplitudes in terms of the above variables requires an additional step. Let us consider, for instance, a three-point amplitude $\cM(\phi^s(p_1), \phi^s(p_2), h(\pH)) $ between two massive spinning states with momenta $p_1$ and $p_2$ and a graviton with momentum $\pH$. The two massive external legs correspond to the incoming and outgoing states of a single classical object, such as a black hole, but they depend on two momenta $p_1$ and $p_2$ and their respective polarisations $\bep_1$ and $\bep_2$. In order to apply eq. \eqref{eq:spindefn} we need to identify a unique momentum $p_1$ describing the classical object, and a unique polarisation $\bep_1$. To do so, we note that $p_2$ is given by a boost\footnote{Note that since our momentum convention is all incoming, this transformation also contains a reflection.}  acting on $p_1$

\begin{equation}
    p_2^\mu = {\Lambda^{\mu}}_{\nu} p_1^\nu = (-\delta^{\mu}_{\nu} + {\omega^{\mu}}_{\nu}) p_1^\nu
 \end{equation} 
with the generator $\omega^{\mu}_{\nu}= \frac{1}{m^2} (p_1^{\mu}\pH_{\nu}-\pH^{\mu}p_{1 \nu})$ as done in refs. \cite{Maybee:2019jus,Arkani-Hamed:2019ymq}. This expression is exact on the three-point kinematics since the generator satisfies $\omega^3 =0$, if $k^2 = p_1 \cdot k =0$, and the quadratic order vanishes as $\omega^2$ is symmetric. For generic higher-point amplitudes, when the massless state is off shell, the generator would require a correction.

Given the generator $\omega$, we can construct the boost in the Weyl representation, $\exp[{-\frac{1}{4}\omega^{\mu \nu} \sigma_{\mu \nu}}]$, such that
 \begin{equation}
\begin{split}
\ket{2^a}&=\ket{1^a} + \frac{1}{2 m_1} \left(\pH \cdot \sigma\right) | 1^a] \\
|2^a] &= -  |1^a] - \frac{1}{2 m_1}\left(\pH \cdot \bar{\sigma}\right)\ket{1^a} ,
\end{split}
\end{equation} where the sign in the second line is generated by the reflection. These Lorentz-boosted spinors appear in ref. \cite{Arkani-Hamed:2019ymq} and are closely related to the on-shell HPET variables in ref. \cite{Aoude:2020onz}.

Note that in general there could be an arbitrary rotation of the little group, but we choose to align the little group of particle $1$ and $2$. This corresponds to identifying $z_2^a = \bar{z}_1^a$, where we use the convention that (un)barred variables correspond to (incoming) outgoing particles. Similarly, we can boost the polarisation of $p_2$ such that
\begin{equation} \label{eq:pol2to1}
    \bep_2^\mu  = \bar{\bep}_1^\mu - \frac{ \pH \cdot \bar{\bep}_1}{m_1^2} \left( p_1^\mu +  \frac{1}{2} \pH^\mu \right).
\end{equation} 

Given these relations we can now express any three-point amplitude in terms of the massive variables $\{p_1,\bar{\bep}_1,\bep_1\}$, together with the massless polarisation $\epH$ and momentum $\pH$, and repackage them into expectation values of the spin operator. The process of converting the amplitude to the spin basis from here can be done in different ways.

For integer spin-$s$ cases, we now can write the amplitude as
\begin{equation}
    \cA(s) = \bar{\bep}_{1 \alpha(s)} {A^{\alpha(s)}}_{\beta(s)}   \bep_{1}^{\beta(s)}
\end{equation} where ${A^{\alpha(s)}}_{\beta(s)}  $ is the function of the momenta $p_1, \pH$ and the massless polarisation $\epH$ that remains after factoring out the massive polarisations. We can decompose the spin-$s$ polarisation vectors directly into spin variables, regardless of the explicit form of ${A^{\alpha(s)}}_{\beta(s)}  $. In the spin-1 case, we have the following decomposition,
\begin{equation} \label{eq:spin1frompols}
    \bar{\bep}_1^\mu \bep_1^\nu = - \angle{S_{(1)}^{(\mu} S_{(1)}^{\nu)}} + \frac{i}{2m} \epsilon^{\mu \nu \rho \sigma} p_{1\rho} \angle{S_{(1)\sigma}} - P_{(1)}^{\mu \nu} 
\end{equation}
where $P_{(1)}^{\mu \nu} = \eta^{\mu \nu} - p_1^\mu p_1^\nu/m^2$ is the spin-1 projector. Similarly, in the spin-2 case we have
\begin{multline}
\label{eq:poltospin2}
\bar{\bep}_1^{\mu_1} \bar{\bep}_1^{\mu_2} \bep_1^{\nu_1} \bep_1^{\nu_2} =
\frac{1}{6} \angle{S_{(2)}^{(\mu_1} S_{(2)}^{\mu_2} S_{(2)}^{\nu_1} S_{(2)}^{\nu_2)}}
- \frac{i}{6 m} {\epsilon^{\mu_1 \nu_1 \kappa}}_\lambda p_{1\kappa} \angle{S_{(2)}^{(\lambda} S_{(2)}^{\mu_2} S_{(2)}^{\nu_2)}}\\
+ \frac{1}{36} \left(   P_{(1)}^{\mu_1 \mu_2} \angle{S_{(2)}^{(\nu_1} S_{(2)}^{\nu_2)}}  + P_{(1)}^{\nu_1 \nu_2} \angle{S_{(2)}^{(\mu_1} S_{(2)}^{\mu_2)}}  + 28 P_{(1)}^{\mu_1 \nu_1} \angle{S_{(2)}^{(\mu_2} S_{(2)}^{\nu_2)}}   \right)\\
- \frac{7 i}{18 m} P_{(1)}^{\mu_1 \nu_1} {\epsilon^{\mu_2 \nu_2 \kappa}}_\lambda p_{1\kappa} \angle{S_{(2)}^{\lambda}}
+ P_{(2)}^{\mu_1 \mu_2 \nu_1 \nu_2} ,
\end{multline}
where we assume symmetrisation in the $\mu_i$ and $\nu_i$ indices separately, and $P_{(2)}^{\mu_1 \mu_2 \nu_1 \nu_2}$ is the spin-2 projector as given in ref. \OurKerr. Similar formulae can be written for any spin, but in general this is quite cumbersome and hides many of the simplifications occurring in the case of on-shell three-point kinematics. Therefore, we choose an alternative approach to tackle large spin amplitudes.

The simplest way to express generic spin-$s$ amplitudes in terms of the spin-$s$ variables is by first converting to spin-$1/2$ building blocks. Any function of the massive spinors $\ket{1^a},|1^a]$ that is homogeneous in $\bar{z}_1$ and $z_1$ can be written as a polynomial in $\angle{S_{(1/2)}}$, for example 
\begin{align} \label{eq:spinfromspinors}
   |\bar{\mathbf{1}}]\bra{\mathbf{1}} &= \left( |1^{(a}]\bra{1^{b)}} +|1^{[a}]\bra{1^{b]}} \right) \bar{z}_{1 a} z_{1 b} \nonumber\\
    &= -m \bar{\sigma} \cdot \angle{S_{(1/2)}}  -\frac{1}{2}  \bar{\sigma} \cdot p_1.
\end{align} 
The next step requires changing the representation of spin-$1/2$ building blocks to generic spin-$s$. The relation for the linear in spin term can be generated from eq. \eqref{eq:spindefn}
\begin{equation} \label{eq:repchangelinear}
\angle{S^\mu_{(s)}} = 2 s \angle{ S^\mu_{(1/2)}  }.
\end{equation} 
However, to generate relations for higher order terms $\Angle{S_{(1/2)}}^n$ one needs to carefully track the little group contractions implied by the angular brackets in $\angle{S^{\mu_1}_{(s)} \dots S^{\mu_n}_{(s)}}$, as given in eq. \eqref{eq:symmexpval}. Generating an identity that holds for arbitrary powers of spin operators in general kinematics is nontrivial and also unnecessary at three points. Indeed, on three-point kinematics, the only independent spin structure that appears is $\angle{\pH \cdot S_{(s)}}$. This simplifies the identity significantly such that we only need the combinatorial factor,
\begin{equation} \label{eq:repchange}
    \Angle{ k \cdot S_{ (1/2) } }^n = \frac{(2s - n)!}{(2s)!} \Angle{ \left(k \cdot S_{(s)}\right)^n }.
\end{equation} This identity reduces to eq. \eqref{eq:repchangelinear} for $n=1$ and is critical in generating the correct classical limit in section \ref{sec:Kerr}.

We will also work with integer-spin amplitudes that are functions of covariant variables. In that case we can express a pair of polarisation vectors in terms of spin-$1/2$ building blocks, and then convert to the appropriate representation using eq. \eqref{eq:repchange}. As an example consider a term that appears later in the leading Regge string amplitude,
\begin{equation}
\begin{split}
    \left(\pH \cdot \bep_2 \pH \cdot \bep_1\right)^n &= 
    \left(\pH \cdot \bar{\bep}_1 \pH \cdot \bep_1\right)^n \\
    &= \left(- 2 \left(\pH \cdot \angle{S_{(1/2)}}\right)^2 \right)^n \\
    &= (-2)^n \frac{(2s - 2 n)!}{(2s)!} \Angle{\left(\pH \cdot S_{(s)}\right)^{2n}}.
\end{split}
\end{equation} In the first step we use eq. \eqref{eq:pol2to1} to relate $\bep_2$ to $\bar{\bep}_1$. Next we use the definition of the polarisation \eqref{eq:poldefn} and eq. \eqref{eq:spinfromspinors} to convert the expression to spin-$1/2$ variables. The final step converts the expression to the generic spin-$s$ representation using eq. \eqref{eq:repchange}, as required for the high-spin limits performed in the next sections.

\subsection{Matching Kerr} \label{sec:Kerr}
In this section we will introduce the classical limit in the context of the Kerr black hole. Since we will reproduce past results it will serve as both a test for the approach discussed above and inspiration for the leading Regge string case discussed in the next sections.

We can define the \textit{classical amplitude} as a contraction of the Kerr energy-momentum tensor with an on-shell graviton state \cite{Vines:2017hyw, Guevara:2018wpp},
\begin{equation}
\label{eq:kerr3pt}
    \ep_{\mu \nu} (k) T^{\mu \nu} (-k)
    = (2\pi)^2 \delta(p_1 \cdot k) \delta(k^2) (\epH \cdot p_1)^2 \exp(k \cdot a),
\end{equation} 
with the graviton polarisation $\ep^{\mu \nu} (k) = \ep_{k}^{\mu} \ep_{k}^{\nu}$. Here $a$ is the \textit{ring radius} of a Kerr black hole, related to the classical spin vector via $a^\mu = S^\mu / m$, and $p_1$ is the black-hole four-momentum, where $p_1 = (m,0,0,0)$ in the rest frame. As discussed by several authors \cite{Guevara:2018wpp,Chung:2018kqs,Guevara:2019fsj,Chung:2019duq,Arkani-Hamed:2019ymq, Haddad:2020tvs,Aoude:2021oqj}, the (positive-helicity) amplitudes that reproduce this result are given by
\begin{equation}
\label{eq:gnima3pt}
    \cM_{\textrm{Kerr}}(\phi^s(p_1), \phi^s(p_2), h^+(\pH))  = (\epH^+ \cdot p_1)^2 \left(\frac{\angle{\bm1 \bm2}}{m}\right)^{2s}.
\end{equation} We will refer to the above as \textit{Kerr amplitudes}, as we are about to see they match eq. \eqref{eq:kerr3pt} in the large-spin limit.

To see this, let us define a prescription to compute the classical limit. Using the procedure discussed previously, any three-point amplitude between two spinning states and a graviton can be expanded in spin variables,
\begin{equation}
\cM (\phi^s(p_1), \phi^s(p_2), h(\pH)) 
= (\epH \cdot p_1)^2
\sum_{n=0}^{2s} c_n^{(s)} \Angle{\kdota{s}{n}} ,
\end{equation}
where we define the operator $a_{(s)}^\mu = S_{(s)}^\mu/m$ for convenience. Given a set of three point amplitudes $\{ \cM(s) \}$ for increasing values of $s$, we define the classical limit as
\begin{equation}
\cM_{cl} \equiv \underset{s \to \infty}{\lim}\cM(\phi^s(p_1), \phi^s(p_2), h(\pH))  = (\epH \cdot p_1)^2
\sum_{n=0}^{\infty} c_n^{(\infty)} (k \cdot a)^n 
\end{equation} 
where the coefficients $c_n^{(\infty)} \equiv \underset{s \to \infty}{\lim} c_n^{(s)}$ are the \textit{spin multipole coefficients} \cite{Vaidya:2014kza}. Note that these are theory-dependent. We have also identified the expectation value $\angle{\kdota{s}{n}}$ with the classical-variable power $(\pH \cdot a)^n$, in the $s \to \infty$ limit. To see why this is sensible, we follow the classical limit prescription defined in ref. \cite{Arkani-Hamed:2019ymq}. We reintroduce $\hbar$ and express the massless momenta in terms of its wavenumber $\bar{k}$ via $k = \hbar \bar{k}$. The classical limit then involves taking $\hbar \to 0$ while $s \to \infty$ such that $\hbar s$ stays finite. This formulation of the limit ensures the combination $\angle{\kdota{s}{n}}$ stays finite and can be identified with the classical quantity $(\pH \cdot a)^n$. 

We can now apply this to the Kerr amplitude in eq. \eqref{eq:gnima3pt}, 
\begin{align}
   \cM_{\textrm{Kerr}}(\phi^{s}(p_1), \phi^{s}(p_2), h^{+}(\pH))
    &= (\epH \cdot p_1)^2 (1 +  \angle{k \cdot a_{(1/2)}})^{2s} \nonumber \\
    &=(\epH \cdot p_1)^2\sum_{n=0}^{2s} \begin{pmatrix} 2s \nonumber \\ n \end{pmatrix} \angle{k \cdot a_{(1/2)}}^n\\
    &= (\epH \cdot p_1)^2 \sum_{n=0}^{2s} \frac{1}{n!} \Angle{\kdota{s}{n}} .
\end{align}
The classical limit requires the high-spin limit, such that $c_n^{(\infty)} = 1/n!$ and $\underset{s \to \infty}{\lim}\cM_{\textrm{Kerr}} \propto  \exp[k \cdot a]$ reproducing the exponential in eq. \eqref{eq:kerr3pt}. However $\cM_{\textrm{Kerr}}$ has the property that all the coefficients $c_n^{(s)}$ are spin independent. This implies that we can read off the classical spin multipole coefficients from any finite spin-$s$ amplitude, up to order $2s$ in the spin vector. This feature is known as \textit{spin universality} \cite{Holstein:2008sw,Holstein:2008sx} and it is a special property of the Kerr amplitudes in eq. \eqref{eq:gnima3pt} \cite{Aoude:2020onz}. As we will see, leading Regge strings do not have such a property and the only way to compute the classical amplitude is via the explicit $s \to \infty$ limit.

Before we move onto the string case, let us briefly discuss the gauge theory case. The amplitudes in eq. \eqref{eq:gnima3pt} can be obtained via a double-copy of three-point amplitudes between two massive spinning particles and a photon \cite{Arkani-Hamed:2017jhn, Johansson:2019dnu}, given by
\begin{equation}
\cA_{\sqrt{\textrm{Kerr}}} (\phi^s(p_1), \phi^s(p_2), A^-(\pH)) = g (\epH^+ \cdot p_1) \left(\frac{\angle{\bm1 \bm2}}{m}\right)^{2s},
\end{equation}
where $g$ is the electromagnetic charge of the spinning particle. We refer to these as the \textit{root-Kerr amplitudes}. Since the spin structure is the same as in $\cM_{\textrm{Kerr}}(s)$, the classical limit is identical to the gravity case, with the exception of the $(\epH^+ \cdot p_1)$ prefactor \cite{Arkani-Hamed:2019ymq}. In particular, for fixed spin $s$, the spin multipole coefficents are $c_{n}^{(\infty)} = c_{n}^{(s)} = 1/n!$, thus the amplitudes exhibit the same spin-universality properties as the Kerr amplitudes.

\section{Leading Regge Superstring Amplitudes}
\label{sect:stringampl}
\subsection{Open String}
We will now consider the three-point amplitudes for the superstring involving two massive states from the leading Regge trajectory and one massless spin-1 boson. We assume the latter is a photon in the rest of this work, since we will connect such amplitudes to electromagnetic currents in section \ref{sect:stringclass}.\footnote{More precisely the string amplitudes considered here come with antisymmetric color factors $f^{abc}$, thus the spin-1 boson should be identified with a gluon of the non-abelian gauge group. However, photon amplitudes can always be obtained from non-abelian ones by projecting to a $U(1)$ subgroup, which has the effect of complexifying the massive matter and setting $f^{abc}=1$.}  A leading Regge state in the open string satisfies the following relation for its mass and spin \cite{green_schwarz_witten_2012,polchinski_1998},
\begin{equation}
\label{eq:leadingregge}
    \alpha' = \frac{s-1}{m^2}.
\end{equation}
If we restrict to integer spin, the relevant three-point amplitude is given in ref. \OliSpin \footnote{The reference provided uses the opposite metric signature and a different overall normalisation. The normalisation in our paper is fixed by matching the spin-$2$ amplitude with the result given in ref.~\cite{Chiodaroli:2021eug} and by requiring the normalisation not to diverge in the $s \to \infty$ limit. This amounts to the following rescaling $\mathcal{A}_3 = g (2\alpha')^{-\frac{1}{2}} \Gamma[s]^{-1} \mathcal{A}_{ref.}$.}
\begin{multline}
\label{eq:openstring3pt}
\mathcal{A}_{3} (\phi^s(p_1), \phi^s(p_2), A(\pH)) =  g  (2 \alpha')^{s} (s-1)!  \sum_{n = 0}^{s} \frac{(- \bep_1 \cdot \bep_2)^n}{(2 \alpha')^n n! [(s-n)!]^2} \times\\
\left(
	- n( \epH \cdot p_1) (-\bep_1 \cdot \pH \, \bep_2 \cdot \pH)^{s-n}
	+ \frac{s(s-n)}{2 \alpha'} \bep_2 \cdot f_\pH \cdot \bep_1 (- \bep_1 \cdot \pH \, \bep_2 \cdot \pH)^{s-n-1}
\right)
\end{multline} 
where $f_{\pH}^{\mu \nu} = 2 \pH^{[\mu} \epH^{\nu]}$ is the linearised field strength for the photon. We have written the amplitude in terms of the on-shell variables for the two massive states, $p_1$, $p_2$, $\bep_1$, $\bep_2$ and the massless state $\pH$, $\epH$. While the superstring amplitudes are naturally defined in $d=10$, we can trivially compactify down to $d=4$ by choosing four-dimensional polarisations, $\bep_i^{d=10} = (\bep_i^{d=4},\bm0)$ where $\bep_i^{d=4}$ were defined in the previous section, and four-dimensional momenta $p_i^{d=10} = (p_i^{d=4},\bm0)$. 

We can also rewrite the amplitude in terms of massive spinor variables,
\begin{multline}
\label{eq:openstringMSH3pt}
\mathcal{A}_{3} (\phi^s(p_1), \phi^s(p_2), A(\pH)) =
-\frac{g (s-1)!}{m^{2s}}(\epH^+ \cdot p_1)   \times \\
\sum_{n = 0}^{s} \frac{(s-1)^{s-n-1}(\mshA \mshS)^n (\mshS-\mshA)^{2s-2n-1} (n(s-1)\mshS-(s^2-n)\mshA)}{n! [(s-n)!]^2} .
\end{multline} 

Following the discussion in the previous section we can convert the amplitude (\ref{eq:openstring3pt}) to the spin basis using the following identities
\begin{equation}
\label{eq:poltospinexamples}
\begin{split}
    &\bep_1 \cdot \bep_2 = -1 + \angle{\pH \cdot a_{(1/2)}}^2, \\
    &\bep_1 \cdot \pH \, \bep_2 \cdot \pH = - 2 m^2 \angle{\pH \cdot a_{(1/2)}}^2, \\
    &\bep_2 \cdot f_{\pH} \cdot \bep_1 = - 2 \epH \cdot p_1 \left( \eta \angle{\pH \cdot a_{(1/2)}} + \angle{\pH \cdot a_{(1/2)}}^2 \right).
\end{split}
\end{equation} 

In the last identity $\eta = \pm 1$ is the helicity of the photon, which we take to be $\eta= 1$ by default. The next step is to use eq. \eqref{eq:repchange} to express the amplitude as a polynomial in spin variables in the appropriate representation. For example, we can generate the amplitude for a spin-2 particle, the lowest-spin massive boson of the open superstring,
\begin{equation}
\mathcal{A}_{3} (\phi^{s=2}(p_1), \phi^{s=2}(p_2), A^{+}(\pH)) =- g  \epH^{+} \cdot p_1
\sum_{n=0}^{4} \frac{1}{n!} \Angle{\kdota{2}{n}}.
\end{equation}
Note that, for this spin, string theory gives the same result as the root-Kerr amplitudes  $\cA_{\sqrt{\textrm{Kerr}}}$ given in section \ref{sec:Kerr}. However, for spin-$3$ states and higher the amplitudes deviates from the truncated exponential form of Kerr already starting from the quadrupole coefficient $c_2^{(s)}$. For a general spin-$s$ boson, the amplitude can be written as the following polynomial in spin,
\begin{equation} \label{eq:finitesopen}
\cA(\phi^s(p_1), \phi^s(p_2), A^{+}(\pH)) 
=-g \epH^{+} \cdot p_1
\sum_{n=0}^{2s} c_n^{(s)} \Angle{\kdota{s}{n}}
\end{equation}
where $c_0^{(s)} = c_1^{(s)} = 1$ and the next four coefficients are:\footnote{The dipole coefficient $c_1^{(s)}$ is proportional to the gyromagnetic ratio, hence its value is universal for string states \cite{Ademollo:1974te}.}
\begin{equation}
\begin{split}
    &c_2^{(s)} = \frac{4s^2-7s+4}{2s(2s-1)}, \qquad \qquad \quad c_3^{(s)} = \frac{2s-3}{2(2s-1)}, \\
    &c_4^{(s)} = \frac{8s^3-32s^2+45s-24}{8s(2s-3)(2s-1)} ,
    \quad c_5^{(s)} = \frac{8 s^2 - 28 s + 23}{24(2s-3)(2s-1)}.
\end{split}
\end{equation}

We have found similar expressions for $c_n^{(s)}$ up to $n = 20$, all rational functions, but we omit them for the sake of simplicity. Clearly these results do not match the root-Kerr case which has coefficients $c_n^{(s)}  = 1/n!$ for any spin state. 

Moreover, all the string multipole coefficients beyond the dipole are spin dependent, hence violating the \textit{spin universality} we discussed in section \ref{sec:Kerr}. Therefore, if we want to extract the classical result, our only option is to take the $s \to \infty$ limit. 
In this limit the coefficient of the $n$'th order multipole is given by
\begin{equation}
    c^{(\infty)}_n= \begin{cases} \frac{1}{(k!)^2} &\textrm{if } n=2k \textrm{ with } k \in \mathbb{N}, \\
    \frac{1}{k!(k+1)!} &\textrm{if }  n=2k+1 \textrm{ with } k \in \mathbb{N}.\end{cases}
\end{equation}

This was computed explicitly up to $n = 20$ and extrapolated to higher $n$ values. The even and odd coefficients are generated by separate modified Bessel functions of the first kind,
\begin{equation}
    I_0 (2 x) = \sum_{k=0}^{\infty} c^{(\infty)}_{2k} x^{2k}, \hspace{10mm}
    I_1 (2 x) = \sum_{k=0}^{\infty} c^{(\infty)}_{2k+1} x^{2k+1}.
\end{equation} Modified Bessel functions are Bessel functions with purely imaginary arguments,  $I_\alpha (x) = i^{\alpha} J_{\alpha}(i x)$.
We can then resum the infinite spin limit of eq. \eqref{eq:finitesopen} to generate the following all-spin result for the classical amplitude
\be
\label{eq:clstringgauge}
\cA_{cl}^{string} \equiv
\lim_{s \to \infty} \cA(\phi^s(p_1), \phi^s(p_2), A^{+}(\pH)) = -g
 \epH^{+} \cdot p_1 \left( 
I_0 (2\pH \cdot a) + I_1(2\pH \cdot a) 
\right).
\ee
As we will show in section \ref{sect:stringclass}, this describes a classical rigid rotating string with a charged endpoint, as expected for classical leading Regge states \cite{zwiebach_2004}.

\subsection{Closed String}
We can repeat the computation for the closed superstring exploiting the simple KLT relation at three points \cite{Kawai:1985xq},
\begin{equation}
\label{eq:closedstring3pt}
\cM (\phi^s(p_1), \phi^s(p_2), h(\pH)) =
\left(\frac{1}{g}\cA(1 \phi^{s/2}, 2 \phi^{s/2}, A(\pH)) \big{|}_{\alpha' \to \alpha'/4} \right)^2.
\end{equation} The rescaling of $\alpha'$ is compensated by the fact that the mass of leading Regge closed string states is given by $m^2 = 4(s/2-1)/\alpha'$. 

The lowest-spin massive state for the leading Regge trajectory of the closed superstring is a spin-4 particle, and its amplitude expanded in spin multipoles is
\begin{equation}
\cM (\phi^{s=4}(p_1), \phi^{s=4}(p_2), h^{+}(\pH)) = ( \epH^{+} \cdot p_1)^2
\sum_{n=0}^{8} \frac{1}{n!} \Angle{\kdota{4}{n}} \, ,
\end{equation}
matching the Kerr amplitude. However, as in the open string, the higher spin states deviate from the Kerr result starting from the quadrupole. The general spin amplitude is
\begin{equation}
\cM(\phi^{s}(p_1), \phi^{s}(p_2), h^{+}(\pH)) 
= ( \epH^{+} \cdot p_1)^2
\sum_{n=0}^{2s} c_n^{(s)} \Angle{\kdota{s}{n}}
\end{equation}
where the first two coefficients are spin-independent $c_0^{(s)} = c_1^{(s)} = 1$. However, the rest of the coefficients have explicit spin dependence, for example the next four coefficients are
\begin{equation}
\begin{split}
    &c_2^{(s)} = \frac{3s^2-7s+8}{2s(2s-1)}, \hspace{5mm} c_3^{(s)} = \frac{3s^2-12s+14}{2(2s-1)(2s-2)},\\
    &c_4^{(s)} = \frac{5s^4-34s^3+91s^2-125s+80}{4s(2s-1)(2s-2)(2s-3)} ,\\
    &c_5^{(s)} = \frac{5s^3-37s^2+97s-95}{12(2s-1)(2s-2)(2s-3)} .
\end{split}
\end{equation} 

As before, we found explicit rational expressions for $c_n^{(s)}$ up to $n = 20$, but we omit them for the sake of simplicity. Taking the $s \to \infty$ limit, the coefficient of the $n$'th order multipole is given by
\begin{equation}
    c^{(\infty)}_n= \begin{cases} \frac{(2k+1)!}{4^{k}(k+1)!(k!)^3} &\textrm{if } n=2k \textrm{ with } k \in \mathbb{N}, \\
    \frac{(2k+1)!}{4^{k}((k+1)!)^2(k!)^2} &\textrm{if }  n=2k+1 \textrm{ with } k \in \mathbb{N}.\end{cases}
\end{equation}

Resumming the above we generate the classical all-spin result for the closed string amplitudes,
\be
\label{eq:closedfinal}
\cM_{cl}^{string} \equiv
\lim_{s \to \infty} \cM (\phi^{s}(p_1), \phi^{s}(p_2), h^{+}(\pH)) = 
(\epH^{+} \cdot p_1)^2 \left( 
I_0 (\pH \cdot a) + I_1(\pH \cdot a) 
\right)^2 .
\ee 
Note that eq. \eqref{eq:closedfinal} is simply the square of the open string result \eqref{eq:clstringgauge} after rescaling $a^\mu \to a^\mu/2$, manifesting the double-copy relation between open and closed strings in the classical limit. This is also true in the case of three-point Kerr amplitudes,

\begin{align} \label{eq:clstringgrav}
\lim_{s \to \infty} \mathcal{M}_{\text{Kerr}} 
&= (\epH \cdot p_1)^2 \exp{\left(\pH \cdot a\right)}
= \left((\epH \cdot p_1) \exp{\left(\frac{\pH \cdot a}{2}\right)} \right)^2 \nonumber \\
&= \left(\lim_{s \to \infty} g^{-1}\mathcal{A}_{\sqrt{\text{Kerr}}} \big{|}_{a^\mu \to a^\mu/2} \right)^2 
\end{align} where we can relate the classical limit of the root-Kerr amplitudes to the Kerr amplitudes after a similar rescaling of the spin. 

We will show in the next section that these classical superstring amplitudes are closely related to the current four-vector/stress-energy tensor of a rigid rotating string coupled to electromagnetism/gravity, such that our results are an instance of a classical double copy between an electromagnetic current and a gravitational stress-energy tensor.

\section{Classical String Solutions}
\label{sect:stringclass}

In this section we study classical string solutions given by rigid strings rotating around their midpoint \cite{Lust1989,zwiebach_2004}. As shown in eq. \eqref{eq:classicalleadingregge}, the angular momentum $J$ and the rest-frame energy $E$ of such solutions are related by the identity $J = \alpha' E^2$. As this matches the relation \eqref{eq:leadingregge} in the large-spin limit, these classical string solutions are classically equivalent to leading Regge string states.\footnote{This requires identifying the rest-frame energy of the classical string, $E$, with the rest-frame energy of the leading Regge particle state, which is just the mass $m$.}
We compute the electromagnetic current and stress-energy tensor for such solutions and, in the spirit of eq. \eqref{eq:kerr3pt}, we use them to define classical amplitudes. These are then shown to match exactly to eqs. \eqref{eq:clstringgauge} and \eqref{eq:closedfinal}, confirming that the classical limit technology we set up in section \ref{sect:recap} produces the expected classical results.

\begin{figure}
    \centering
    \includegraphics[width=11cm]{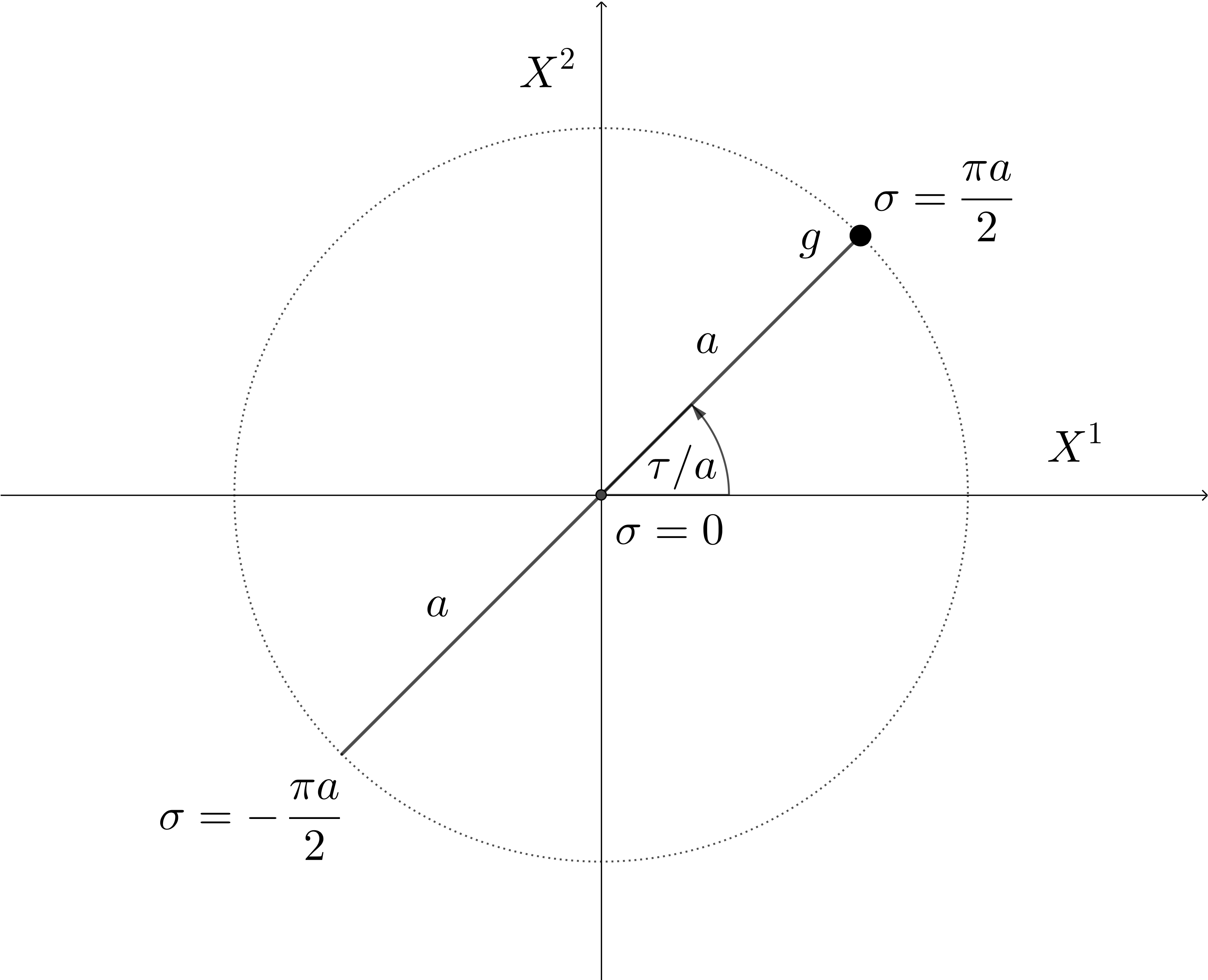}
    \caption{Rigid open string of length $2 a$ with a charge $g$ at one endpoint rotating around its centre point with angular velocity $1/a$.} 
    \label{fig:rigidstring_charged}
\end{figure}
\subsection{Leading Regge Open String Solution}
The classical open string solution that corresponds to leading Regge states is a rigid string rotating around its midpoint. Since we need the string to couple to a photon, we place a charge $g$ at one endpoint, as in figure \ref{fig:rigidstring_charged}. 

We can treat this system using classical electrodynamics and construct the current four-vector from the charge density $\rho$ and the current $\mathbf{j}$,
\begin{align}
j^\mu(x) &= \left(\rho(x), \mathbf{j}(x) \right)^\mu =\frac{g}{a}  \delta(r - a)\delta(\phi - t/a) \delta(z) n^\mu,
\end{align} with $n^\mu = (1, \hat{\phi}) = (1, - \sin(t/a), \cos( t/a),0)$. Note that we work with $c=1$. In appendix \ref{sec:openstringdetails} we provide a first-principles derivation of $j^\mu$. 

The current four-vector is analogous to the energy-momentum tensor in the gravity case, in the sense that it describes the linear coupling to the massless gauge field. More precisely, the interaction term in the Lagrangian is proportional to $h_{\mu \nu} T^{\mu \nu}$ in the gravity case and $A_{\mu} j^\mu$ in electromagnetism. Therefore we can construct a classical three-point amplitude in the same way as eq. \eqref{eq:kerr3pt}, by contracting $j^\mu$ with an on-shell photon polarisation and Fourier-transforming to momentum space.

In order to construct an appropriate polarisation vector, $\epH$, let us first consider the constraints dictated by the three-point kinematics. On these kinematics the wave number vector $k^\mu$ must satisfy
\begin{equation}
\label{eq:onshell3pt}
    k^2 = p_1 \cdot k = \epH \cdot k = 0,
\end{equation} 
where in this case $p_1$ is the total four-momentum of the string. We choose to work in the centre of mass frame, where the total energy of the string is $E$ and the four-momentum is $p_1 = (E, \mathbf{0})$. In order to satisfy eq. \eqref{eq:onshell3pt} we need complexified kinematics, a suitable choice is 
\begin{equation}
    k^\mu = E_\gamma (0, i,0,-1), \qquad     \epH^\mu = \frac{1}{\sqrt{2}}(1,0,-1,0).
\end{equation} Given this choice we can proceed with the Fourier transform of $\epH \cdot j$, 
\begin{equation}
    \begin{split}
\mathcal{FT}[\epH \cdot j] &= \frac{1}{2 \pi a}\int d^4 x \: e^{i k \cdot x} \: \frac{g}{a} \: \epH \cdot n \: \delta(r - a)\delta(\phi - t/a) \delta(z) \\
&= \frac{g }{2 \pi \sqrt{2}} \int_{-\pi}^{\pi}  d \phi \: e^{ a E_\gamma \cos \phi} \left(1 + \cos \phi \right) 
\\
&=  \frac{ g}{\sqrt{2}} \left[ I_0( a E_\gamma)+ I_1( a E_\gamma ) \right].
\end{split}
\end{equation} Where we used a well known integral representation for the modified Bessel functions.
Compare this result to the classical limit of the string amplitude \eqref{eq:clstringgauge} given in the previous section. While the same functions appear in both expressions, the prefactor and the arguments are written in different variables.

The rotating string is an extended object with only orbital angular momentum $J^i =(0,0, J_z)$. Meanwhile, the amplitude result in \eqref{eq:clstringgauge} repackages the string as an effective point particle, with mass $m$ and intrinsic spin, $\Angle{S^\mu} = (0,0,0,S_z)$. The total energy of the effective point particle, $m$ in the centre-of-mass frame, corresponds to the total energy of the string, $E$, such that we can identify $E =m$. Likewise we can identify the spin of the effective point particle with the total angular momentum of the string, $S_z = J_z$. The total energy and total angular momentum of the string are calculated in the appendix \ref{sec:openstringdetails}, 
\begin{equation}
    E = \frac{a}{2 \alpha'}\,, \hspace{8mm}J_z = \frac{a^2}{4 \alpha'}.
\end{equation} 

Using the above and recalling that $\pH =E_\gamma (0, i,0,-1)$, we can confirm that the arguments of the Bessel functions match,
\begin{equation}
    \frac{2 \pH \cdot S}{m} = \frac{4 E_\gamma \alpha' }{a} S_{z} = a E_{\gamma}.
\end{equation}

The next step is to consider the coefficient of the Bessel functions. The normalisation of the amplitude was chosen to be finite in the classical limit and it is given by $p_1 \cdot \epH$. In the frame specified above this reduces to the mass of the point particle $m$.

Thus we have the match between the classical open string amplitude, $\mathcal{A}_{cl.}$, computed in section  \ref{sect:stringampl}, and the classical solution from first principles,
\begin{equation}
 \cA_{cl}^{string} = m \mathcal{FT}[\epH \cdot j] = \frac{m g}{\sqrt{2}} \left[ I_0\left( a E_\gamma\right)+ I_1\left( a E_\gamma \right) \right].
\end{equation}

\subsection{Leading Regge Closed String Solution}

We now move to the classical closed string configuration. The leading Regge solution now corresponds to a folded rigid string rotating around its centre point, in the same setup as figure \ref{fig:rigidstring_charged}. This configuration can be represented by the following stress-energy tensor
\begin{align}
    T^{\mu \nu} &= \frac{1}{\pi \alpha'} \frac{a\gamma(r)}{r}  n^{\mu \nu}(r,\phi)\delta(z) \left[\delta(t- a\phi) + \delta(t - a\phi - a \pi)\right] \Theta(a-r), 
\end{align} with the Lorentz factor, $\gamma(r) = \left(1 - \frac{r^2}{a^2}\right)^{-1/2}$, and the matrix
\begin{equation}
 n^{\mu \nu}(r,\phi) =   \begin{pmatrix}
 1 & - \frac{r}{a} \sin{\phi} & \frac{r}{a} \cos{\phi} & 0 \\ 
 - \frac{r}{a} \sin{\phi} & \frac{r^2}{a^2} - \cos^2{\phi} & -\frac{1}{2} \sin{2 \phi} & 0\\
 \frac{r}{a} \cos{\phi} & -\frac{1}{2} \sin{2\phi} & \frac{r^2}{a^2} - \sin^2{\phi} & 0 \\
 0 & 0 & 0 & 0
 \end{pmatrix}.    
\end{equation} 
The derivation of this result from the classical string solution is given in appendix \ref{sec:openstringdetails}, with an explicit check that $\nabla_\mu T^{\mu \nu}=0$. 

Now we can follow the same process as in the open string case, first we saturate the Lorentz indices of $T^{\mu \nu}$ with the massless polarisation vectors, and then we take the Fourier transform,
\begin{align}
    \mathcal{FT}[\epH \cdot T \cdot \epH] &=\frac{1}{2 \pi^2 \alpha' } \! \int \! \! d^4x e^{i k \cdot x} \frac{\gamma(r)}{r}  \epH \! \cdot \! n \! \cdot \! \epH \, \delta(z) \left(\delta(t\!-\!a\phi) + \delta(t\!-\!a\phi\!-\!a\pi)\right) \Theta(a\!-\!r) \nonumber\\
    &=  \frac{1}{2 \pi^2  \alpha'} \! \int dr d \phi \gamma(r) \left( \frac{r}{a} + \cos{\phi} \right)^2 e^{r E_\gamma \cos{\phi}} \Theta(a\!-\!r)\nonumber \\
    &= \frac{m}{2} \left[ I_{0}\left(\frac{a E_\gamma}{2}\right) + I_{1}\left(\frac{a E_\gamma}{2}\right) \right]^2
\end{align}  
where we used the closed-string relation $m = E = a/\alpha'$. In order to compare this result to eq. \eqref{eq:clstringgrav} we need to proceed in a similar manner to the open string case. Note that for closed strings we have $E_{\textrm{closed}} = 2 E_{\textrm{open}}$ and $J_{\textrm{closed}}=  2 J_{\textrm{open}}$, see appendix \ref{sec:openstringdetails}. However the ratio $J/E$ is invariant, such that we have the same relations as in the open string case. Hence the Bessel function arguments and the prefactor in the amplitude can be expressed as
\begin{equation}
    \frac{\pH \cdot S}{m} = \frac{a E_\gamma}{2}, \qquad (p_1 \cdot \epH)^2 =\frac{m^2}{2}.
\end{equation} 

Thus we have the match between the classical closed string amplitude, $\mathcal{M}_{cl.}$, computed in section \ref{sect:stringampl} and the classical solution from first principles,
\begin{equation}
 \cM_{cl}^{string} = m \mathcal{FT}[\epH \cdot T \cdot \epH] =\frac{ m^2 }{2} \left[ I_0\left(\frac{a E_\gamma}{2}\right)+ I_1\left(\frac{a E_\gamma}{2} \right) \right]^2.
\end{equation}

\section{Conclusion}
\label{sect:conclusion}
In this work, we computed the classical limit of superstring amplitudes involving two massive leading Regge states and a photon or graviton. In the process, we set up a prescription for the classical limit of three-point amplitudes and discussed the importance of the spin to infinity limit to reproduce the classical spin-multipole coefficients. We found that the gauge theory and gravity results matched the electromagnetic current and stress-energy tensor obtained from fully-classical string solutions, and we highlighted the classical double copy relation between the two.

We found in section \ref{sect:stringampl} that generic amplitudes do not obey the \textit{spin universality} observed in the Kerr case, as shown explicitly in the string case. In particular, any amplitude for two massive states of finite spin $s$ and a photon or graviton can be expanded as a polynomial of the normalised spin operator $a^\mu = S^\mu/m$,
\begin{equation}
 \cM(s) \propto \sum_{n=0}^{2s} c_n^ {(s)} \Angle{\kdota{s}{n}}, \nonumber
\end{equation} such that, in general, the coefficients $c_n^{(s)}$ depend on the spin quantum number $s$. For example, the coefficient of the spin-quadrupole for the open string had the following functional dependence on $s$: $c_{2}^{(s)}= \frac{4s^2-7s+4}{2s(2s-1)}$. 

This feature indicates that, in general, one cannot read off the correct classical spin multipole coefficients from finite-spin quantum amplitudes. Instead the classical coefficients correspond to the $s \to \infty$ limit of these fixed spin coefficients. Applying this approach to the leading Regge open and closed superstring amplitudes yields the following all-spin classical results,
\begin{align} \label{eq:results}
    \lim_{s \to \infty} \cA(\phi^s(p_1),\phi^s(p_2),A^{+}(k)) &= g \, \epH^{+} \cdot p_1 \left[ 
I_0 (2\pH \cdot a) + I_1(2\pH \cdot a) 
\right] , \nonumber \\
\lim_{s \to \infty} \cM (\phi^s(p_1), \phi^s(p_2), h^{+}(k)) &= 
(\epH^{+} \cdot p_1)^2 \left[ 
I_0 (\pH \cdot a) + I_1(\pH \cdot a) 
\right]^2. \nonumber
\end{align} Clearly these leading Regge strings do not reproduce the exponential that appears in the Kerr stress-energy tensor, instead there are modified Bessel functions appearing. We noted that the classical result inherits the double copy structure that the finite-spin string amplitudes admit at three points. This feature is also present in the Kerr case, as shown in eq. \eqref{eq:clstringgrav}.

We were able to verify the results above by matching to purely classical calculations. The configuration associated to classical leading Regge states is known to be that of a rotating rigid string \cite{zwiebach_2004}. The open string corresponds to a string with a charged endpoint, while the closed string corresponds to a folded string coupled to gravity. We constructed the current four-vector and stress-energy tensor in the respective cases and, as was done for the Kerr case, transformed it to momentum space and contracted it with on-shell polarisation vectors corresponding to the photon/graviton. This procedure yields a classical three-point amplitude that can be compared to the infinite-spin limit results. The Bessel function structure falls out immediately and the arguments can be matched in the rest frame of the particle.

In this paper we only considered leading Regge states but one could naturally extend the analysis to other string states. It is possible that the amplitudes for a subleading Regge trajectory agree with Kerr in the spin to infinity limit. Another possibility would be to consider a superposition of string states, such as a string coherent state, and use the methods outlined in this work to try and match black-hole observables. If black-hole amplitudes were to be found within string theory, the classical limits of higher point string amplitudes could shed some light on the high-spin Compton amplitude for Kerr.

As string theory provides a consistent theory of massive interacting higher-spin particles, it was a natural setting to explore classical limits that require spin to infinity limits. However there are alternative approaches to building consistent theories of spinning particles, which have emerged from the higher-spin research community. A feasible future direction could be to explore the relation between the Kerr amplitudes and the higher-spin amplitudes that such alternative theories provide, as initiated in ref. \cite{Chiodaroli:2021eug}.

\acknowledgments

We especially thank Henrik Johansson, Kays Haddad and Paolo Di Vecchia for their contributions at early stages of this project, and many helpful discussions since. We would also like to thank Francesco Alessio, Maor Ben-Shahar, Chen Huang, Alexander Ochirov, Rodolfo Russo, Oliver Schlotterer and Justin Vines for enlightening discussions related to this work. We thank KITP for their hospitality during our stay at the “High-Precision Gravitational Waves” program, where part of this work was completed. This research was supported in part by the National Science Foundation under Grant No. NSF PHY-1748958. This research is supported in part by the Knut and Alice Wallenberg Foundation under grants KAW 2018.0116 (From Scattering Amplitudes to Gravitational Waves) and KAW 2018.0162 (Exploring a Web of Gravitational Theories through Gauge-Theory Methods), the Swedish Research Council under grant 621-2014-5722, and the Ragnar Söderberg Foundation (Swedish Foundations’ Starting Grant).

\appendix
\section{Deriving the Classical String Solutions} \label{sec:openstringdetails}

Let us consider a rigid open string of uniform mass density, rotating around its centre point in the same setup as figure \ref{fig:rigidstring_charged}. This configuration can be represented by the following string solution \cite{Ademollo:1974te,Lust1989,zwiebach_2004}:
\begin{align}
    &X^0 = \tau ,\nonumber\\
    &X^1 = a \cos{\frac{\tau}{a}} \sin{\frac{\sigma}{a}} ,\nonumber\\
    &X^2 = a \sin{\frac{\tau}{a}} \sin{\frac{\sigma}{a}} ,\nonumber\\
    &X^3 = 0 ,
\end{align} with $\sigma \in [-\frac{\pi a}{2}, \frac{\pi a}{2}]$ and $\tau \in (- \infty, \infty)$.\footnote{It can be shown that this configuration solves the string theory equations of motion in conformal gauge. We refer the reader to the references provided for more detail.} The extension to the closed string is simply achieved by extending the range of $\sigma$ to $[-\pi a, \pi a]$ and it describes a rigid folded string rotating around its centre point in the same fashion. It is convenient to express this in polar coordinates; in order to keep $r>0$ we define different coordinates on the two branches of the string. The corresponding polar coordinates for the open string are
\begin{center}
\begin{tabular}{c|c|c}
  & $\sigma \in \left[0, \frac{\pi a}{2} \right]$  &  $ \sigma \in \left[-\frac{\pi a}{2},0\right]$ \\ \hline
   $r'$  &  $a \sin \frac{\sigma}{a}$ & $-a \sin \frac{\sigma}{a}$ \\
   $\phi'$ & $\frac{\tau}{a}$ & $\frac{\tau}{a} - \pi$
\end{tabular}
\end{center}  while the closed string requires the definitions on the additional domains
\begin{center}
\begin{tabular}{c|c|c}
  & $\sigma \in  [\frac{\pi a}{2}, \pi a]$  &  $ \sigma \in [- \pi a, -\frac{\pi a}{2}]$ \\ \hline
   $r'$  &  $a \sin \left(\pi - \frac{\sigma}{a}\right)$ & $ a \sin\left(\pi + \frac{\sigma}{a} \right)$ \\
   $\phi'$ & $\frac{\tau}{a}$ & $\frac{\tau}{a} - \pi$
\end{tabular}
\end{center}
Below we study the coupling of this solution to the electromagnetic and gravitational fields. 

\subsection{Coupling to Gravity}

We consider the string action in a curved background
\begin{equation}
    S 
    = \frac{1}{4 \pi \alpha'} \int d^4 y d^2 \sigma  \delta^4(y-X(\sigma)) \partial_a X^\mu \partial^a X^\nu g_{\mu \nu} (y),
\end{equation}
and hence we can compute the linearised energy-momentum tensor
\begin{equation}
    T^{\mu \nu} 
    = \frac{2}{\sqrt{-g}} \frac{\delta S}{\delta g_{\mu \nu}(y)} \Big{|}_{g = \eta}
    = \frac{1}{2 \pi \alpha'} \int d^2 \sigma \delta^4(y-X(\sigma)) \partial_a X^\mu \partial^a X^\nu,
\end{equation}
where
\begin{align}
    &\partial_a X^\mu \partial^a X^\nu =
    \begin{pmatrix}
    1 & - S_\tau S_\sigma & C_\tau S_\sigma & 0 \\
    - S_\tau S_\sigma & S_\tau^2 S_\sigma^2 - C_\tau^2 C_\sigma^2 & - S_\tau C_\tau & 0 \\ 
    C_\tau S_\sigma & - S_\tau C_\tau & C_\tau^2 S_\sigma^2 - S_\tau^2 C_\sigma^2 & 0\\
    0 & 0 & 0 & 0
    \end{pmatrix} , \nonumber \\
    &\delta^4(y-X(\sigma)) 
    = \delta(t-\tau) \delta(z) \delta(x-a C_\tau S_\sigma) \delta(y-a S_\tau S_\sigma)
\end{align} and we have used $C_\alpha \equiv \cos{\frac{\alpha}{a}}, S_\alpha \equiv \sin{\frac{\alpha}{a}}$ and $y^\mu = (t,x,y,z)$.

Now let us integrate out the worldsheet coordinates $\{\sigma,\tau\}$. The form of $T^{\mu\nu}$ above suggests doing the integral in polar coordinates, which will introduce multiple Jacobian factors. As well as the measure, the delta function $\delta^4(y-X(\sigma)) $ will pick up a Jacobian factor due to the change of variables,
\begin{align}
    \delta^4(y-X(\sigma)) &= \delta(t-\tau) \delta(z) \delta(x-X^1) \delta(y-X^2) \nonumber \\
    &= \frac{1}{|\mathcal{J}(r', \phi'; X
    ^1,X^2)|} \delta(t-\tau) \delta(z) \delta(r-r') \delta(\phi - \phi')
\end{align} where $X^1 = r' \cos \phi'$, $X^2 = r' \sin \phi'$. The Jacobian factors needed are 
\begin{align}
&|\mathcal{J}(r', \phi';X^1,X^2)| = r', \nonumber \\
&|\mathcal{J}(r', \phi'; \sigma, \tau)| =  a \gamma(r') \,
\end{align} where $\gamma(r') = \left(1 - \frac{{r'}^2}{a^2}\right)^{-1/2}$ is the Lorentz factor. Thus the integral can be simplified to
\begin{multline}
    T^{\mu \nu}  = \frac{1}{2 \pi \alpha'}  \int dr'  d \phi' \frac{|\mathcal{J}(r', \phi'; \sigma, \tau)|}{|\mathcal{J}(r', \phi'; x',y')|} \left[ \delta(t-a \phi') + \delta(t-a \phi' - a \pi)  \right] \\
\times \delta(z) \delta(r-r') \delta(\phi - \phi') \partial_a X^\mu \partial^a X^\nu  .
\end{multline}
Performing the integral we get the expression
\begin{equation}
    T^{\mu \nu} = \frac{N_{\text{string}}}{2 \pi \alpha'} \frac{a \gamma(r)}{r} \left[ \delta(t- a \phi) + \delta(t- a \phi - a \pi)\right] \Theta(a-r) \delta(z) n^{\mu \nu} 
    ,
\end{equation} where 
\begin{equation}
 n^{\mu \nu}(r,\phi) = \begin{pmatrix}
 1 & - \frac{r}{a} \sin{\phi} & \frac{r}{a} \cos{\phi} & 0 \\ 
 - \frac{r}{a} \sin{\phi} & \frac{r^2}{a^2} - \cos^2{\phi} & -\frac{1}{2} \sin{2 \phi} & 0\\
 \frac{r}{a} \cos{\phi} & -\frac{1}{2} \sin{2\phi} & \frac{r^2}{a^2} - \sin^2{\phi} & 0 \\
 0 & 0 & 0 & 0
 \end{pmatrix}
\end{equation} and we have introduced the normalisation $N_{\text{string}}$ which is equal to $1$ in the open case and $2$ in the closed case, since the enlarged $\sigma$ domain effectively corresponds to summing two open strings together.

Given $T^{\mu \nu}$ we can now calculate the total energy $E$,
\begin{align}
E &= \int d^3 x T^{tt} \nonumber\\
&=  \frac{N_{\text{string}}}{2 \pi \alpha'}  \int r dr d\phi dz \frac{a \gamma(r)}{r} \left[ \delta(t- a \phi) + \delta(t- a \phi - a \pi)\right]\Theta(a-r) \delta(z)  \nonumber \\
&=  \frac{N_{\text{string}}}{\pi \alpha'}  \int_0^a r dr \frac{ \gamma(r)}{r}  = N_{\text{string}} \frac{a}{2 \alpha'} .
\end{align}
We can also compute the total angular momentum $J$. Since the string is rotating in the $z=0$ plane the angular momentum is only in the $z$ direction, $J= J_z$, 
\begin{align}
J &= \int d^3 x \left( x T^{ty} - y T^{tx} \right) \nonumber\\
&=  \frac{N_{\text{string}}}{2 \pi \alpha'}  \int r dr d\phi dz \left( r \gamma(r) \right)  \left[ \delta(t- a \phi) + \delta(t- a \phi - a \pi)\right] \Theta(a-r) \delta(z) \nonumber\\
&=  \frac{N_{\text{string}}}{2\pi \alpha'} \int_0^a  dr \frac{r^2 \gamma(r)}{a} = N_{\text{string}} \frac{a^2}{4 \alpha'}.
\end{align} 
The results for $E$ and $J$ above can be used to recover the leading Regge trajectory relation at the classical level. For instance for the open string we have
\begin{equation}
\label{eq:classicalleadingregge}
    J = \alpha' E^2
\end{equation} 
which is analogous to eq. \eqref{eq:leadingregge} in the large spin limit, suggesting that this classical string configuration indeed corresponds to the classical limit of leading Regge string amplitudes. 

Another important check is that this energy-momentum tensor in conserved, $\nabla_\mu T^{\mu \nu}=0$. This is easier to show if we convert the tensor from Cartesian coordinates $x^\mu = (t,x,y,z)$ to cylindrical polar coordinates $x'^\mu = (t,r,\phi,z)$,
\begin{equation}
n^{\mu \nu}_{pol} 
= \frac{\partial x'^{\mu} }{\partial x^{\alpha}} \frac{\partial x'^{\nu} }{\partial x^{\beta}} n^{\alpha \beta} \\
= \begin{pmatrix}
 1 & 0 & \frac{1}{a} & 0 \\ 
 0 &-1 +\frac{r^2}{a^2} &0 & 0\\
 \frac{1}{a} & 0 & \frac{1}{a^2} & 0 \\
 0 & 0 & 0 & 0
\end{pmatrix}. 
\end{equation}
Note that in this set of coordinates we have some non-vanishing Christoffel symbols, namely $\Gamma^r_{\phi \phi} = -r$ and $\Gamma^\phi_{r \phi} = \Gamma^\phi_{\phi r} = 1/r$. Hence we get,
\begin{equation} \label{}
    \int d^4 x f(x) \nabla_\mu T^{\mu \nu} 
    = \int d^4 x f(x)
    \begin{pmatrix}
    \partial_t T^{t t} + \partial_\phi T^{\phi t}\\
    \partial_r T^{r r} + \frac{1}{r} T^{r r} - r T^{\phi \phi}\\
    \partial_t T^{t \phi} + \partial_\phi T^{\phi \phi}\\
    0
    \end{pmatrix} .
\end{equation}
Let us begin with the $t$ component. We get
\begin{multline} \label{eq:timeconservation}
    \int d^4 x f(x) \nabla_\mu T^{\mu t} \\
    = \frac{a N_{\text{string}}}{2 \pi \alpha'} \int d^4 x f(x) \frac{\gamma(r)}{r} \delta(z)\Theta(a-r) \left( \partial_t \delta(t-a\phi) + \frac{1}{a} \partial_\phi \delta(t-a\phi) \right) .
\end{multline}
This is equal to zero due to the identity $\partial_\phi \delta(t-a\phi) = - a \partial_t \delta(t-a\phi)$. Note that we have omitted the term proportional to $\delta(t-a\phi-a\pi)$ for convenience, but it can be obtained from the above by shifting $\phi \to \phi + \pi$ and the same argument applies. Next we consider the $\phi$ component,
\begin{multline} \label{}
    \int d^4 x f(x) \nabla_\mu T^{\mu \phi} \\
    = \frac{a N_{\text{string}}}{2 \pi \alpha'} \int d^4 x f(x) \frac{\gamma(r)}{r} \delta(z)\Theta(a-r) \left( \frac{1}{a} \partial_t \delta(t-a\phi) + \frac{1}{a^2} \partial_\phi \delta(t-a\phi) \right) .
\end{multline}
This is just a rescaling of eq. \eqref{eq:timeconservation}, and hence it vanishes for the same reason. The $r$ component is slightly more involved, and it is given by
\begin{align} \label{}
    \int d^4 x f(x) T^{\mu r} 
    &= \frac{a N_{\text{string}}}{2 \pi \alpha'} \! \int d^4 x f(x) \delta(t\!-\!a\phi) \delta(z)
    \left\{ -\partial_r \left(\frac{\Theta(a\!-\!r)}{r \gamma(r)}\right) - \Theta(a\!-\!r) \frac{\gamma(r)}{r^2} \right\} \nonumber \\
    &= \frac{a N_{\text{string}}}{2 \pi \alpha'} \! \int d^4 x f(x) \delta(t\!-\!a\phi) \delta(z)
    \frac{1}{r \gamma(r)} \delta(a\!-\!r)
\end{align}
where we have used $\partial_r \Theta(a-r) = -\delta(a-r)$. Since $(\gamma(a))^{-1} = 0$, the integrand vanishes on the support of the delta function $\delta(a-r)$. As before, the same argument applies for the term proportional to $\delta(t-a\phi-a\pi)$. Since all components are zero, the identity $\nabla_\mu T^{\mu \nu} = 0$ is verified.

\subsection{Coupling to Electromagnetism}
The coupling of an open string to the electromagnetic field is given by the following action \cite{Tong:2009np},
\begin{equation}
     S_{EM}=i g \int d^2 \sigma \delta \left(\sigma - \frac{\pi a}{2}\right) A^{a}_{\mu}(X) \dot{X}^\mu 
\end{equation}
where $g$ is the charge. Note that we chose to place it only at one endpoint, $\sigma = \pi a/2$, but in principle we could add a similar term at $\sigma = -\pi a/2$. The string solution gives
\begin{equation}
\begin{split}
    &X^\mu(\tau,\pi a/2) = \left( \tau,a\cos{\frac{\tau}{a}},a\sin{\frac{\tau}{a}},0 \right),\\
    &\dot{X}^\mu(\tau,\pi a/2) = \left( 1,-\sin{\frac{\tau}{a}},\cos{\frac{\tau}{a}},0 \right) \equiv n^\mu (\tau/a) .
\end{split}
\end{equation}
From this we can derive the electromagnetic current,
\begin{align}
    j^\mu (y) &= -i\frac{\partial S_{EM}}{\partial A_{\mu}(y)} 
    = g \int d\tau \, n^\mu(\tau/a) \delta(\tau-t) \delta\left(x-a\cos{\frac{\tau}{a}}\right)\delta\left(y-a\sin{\frac{\tau}{a}}\right) \delta(z) \nonumber\\
    &= g \, n^\mu(t/a) \delta\left(x-a\cos{\frac{t}{a}}\right)\delta\left(y-a\sin{\frac{t}{a}}\right) \delta(z) \nonumber\\
    &= \frac{g}{a} n^\mu(\phi) \delta(r-a)\delta\left(\phi-\frac{t}{a}\right) \delta(z)
\end{align} where we have chosen the normalisation such that $\int d^3 x j^0(x) = g$.
Similarly to the closed string case, we can check the conservation condition $\nabla_\mu j^\mu = 0$ by integrating it against an arbitrary function $f(x)$. It is again convenient to rewrite $n^\mu$ in cylindrical polar coordinates,
\begin{equation}
    n^\mu_{pol} = \left(1,0,\frac{1}{r},0\right) .
\end{equation}
Hence we have
\begin{equation}
\int d^4 x f(x) \nabla_\mu j^\mu 
= \frac{c g}{a}\int d^4 x f(x) \delta(r - a) \delta(z)
\left\{ \partial_t \delta \left(\phi - \frac{t}{a}\right) + \frac{1}{a} \partial_\phi \delta \left(\phi - \frac{t}{a}\right) \right\}
\end{equation}
where in this case $\nabla_\mu j^\mu = \partial_\mu j^\mu$ since there are no non-zero Christoffel symbols contributing. This integral vanishes because of the identity $\partial_\phi \delta (\phi - t/a) = - a \partial_t \delta (\phi - t/a)$, which we used already in eq. \eqref{eq:timeconservation}.

\bibliographystyle{JHEP}
\bibliography{references}

\providecommand{\href}[2]{#2}\begingroup\raggedright\begin{thebibliography}{100}

\bibitem{Vaidya:2014kza}
V.~Vaidya, \emph{{Gravitational spin Hamiltonians from the S matrix}},
  \href{https://doi.org/10.1103/PhysRevD.91.024017}{\emph{Phys. Rev. D}
  {\bfseries 91} (2015) 024017}
  [\href{https://arxiv.org/abs/1410.5348}{{\ttfamily 1410.5348}}].

\bibitem{Arkani-Hamed:2017jhn}
N.~Arkani-Hamed, T.-C.~Huang and Y.-t.~Huang, \emph{{Scattering amplitudes for
  all masses and spins}},
  \href{https://doi.org/10.1007/JHEP11(2021)070}{\emph{JHEP} {\bfseries 11}
  (2021) 070} [\href{https://arxiv.org/abs/1709.04891}{{\ttfamily
  1709.04891}}].

\bibitem{Vines:2017hyw}
J.~Vines, \emph{{Scattering of two spinning black holes in post-Minkowskian
  gravity, to all orders in spin, and effective-one-body mappings}},
  \href{https://doi.org/10.1088/1361-6382/aaa3a8}{\emph{Class. Quant. Grav.}
  {\bfseries 35} (2018) 084002}
  [\href{https://arxiv.org/abs/1709.06016}{{\ttfamily 1709.06016}}].

\bibitem{Guevara:2017csg}
A.~Guevara, \emph{{Holomorphic Classical Limit for Spin Effects in
  Gravitational and Electromagnetic Scattering}},
  \href{https://doi.org/10.1007/JHEP04(2019)033}{\emph{JHEP} {\bfseries 04}
  (2019) 033} [\href{https://arxiv.org/abs/1706.02314}{{\ttfamily
  1706.02314}}].

\bibitem{Guevara:2018wpp}
A.~Guevara, A.~Ochirov and J.~Vines, \emph{{Scattering of Spinning Black Holes
  from Exponentiated Soft Factors}},
  \href{https://doi.org/10.1007/JHEP09(2019)056}{\emph{JHEP} {\bfseries 09}
  (2019) 056} [\href{https://arxiv.org/abs/1812.06895}{{\ttfamily
  1812.06895}}].

\bibitem{Chung:2018kqs}
M.-Z.~Chung, Y.-T.~Huang, J.-W.~Kim and S.~Lee, \emph{{The simplest massive
  S-matrix: from minimal coupling to Black Holes}},
  \href{https://doi.org/10.1007/JHEP04(2019)156}{\emph{JHEP} {\bfseries 04}
  (2019) 156} [\href{https://arxiv.org/abs/1812.08752}{{\ttfamily
  1812.08752}}].

\bibitem{Guevara:2019fsj}
A.~Guevara, A.~Ochirov and J.~Vines, \emph{{Black-hole scattering with general
  spin directions from minimal-coupling amplitudes}},
  \href{https://doi.org/10.1103/PhysRevD.100.104024}{\emph{Phys. Rev. D}
  {\bfseries 100} (2019) 104024}
  [\href{https://arxiv.org/abs/1906.10071}{{\ttfamily 1906.10071}}].

\bibitem{Chung:2020rrz}
M.-Z.~Chung, Y.-t.~Huang, J.-W.~Kim and S.~Lee, \emph{{Complete Hamiltonian for
  spinning binary systems at first post-Minkowskian order}},
  \href{https://doi.org/10.1007/JHEP05(2020)105}{\emph{JHEP} {\bfseries 05}
  (2020) 105} [\href{https://arxiv.org/abs/2003.06600}{{\ttfamily
  2003.06600}}].

\bibitem{Kawai:1985xq}
H.~Kawai, D.C.~Lewellen and S.H.H.~Tye, \emph{{A Relation Between Tree
  Amplitudes of Closed and Open Strings}},
  \href{https://doi.org/10.1016/0550-3213(86)90362-7}{\emph{Nucl. Phys. B}
  {\bfseries 269} (1986) 1}.

\bibitem{Bern:2008qj}
Z.~Bern, J.J.M.~Carrasco and H.~Johansson, \emph{{New Relations for
  Gauge-Theory Amplitudes}},
  \href{https://doi.org/10.1103/PhysRevD.78.085011}{\emph{Phys. Rev. D}
  {\bfseries 78} (2008) 085011}
  [\href{https://arxiv.org/abs/0805.3993}{{\ttfamily 0805.3993}}].

\bibitem{Bern:2010ue}
Z.~Bern, J.J.M.~Carrasco and H.~Johansson, \emph{{Perturbative Quantum Gravity
  as a Double Copy of Gauge Theory}},
  \href{https://doi.org/10.1103/PhysRevLett.105.061602}{\emph{Phys. Rev. Lett.}
  {\bfseries 105} (2010) 061602}
  [\href{https://arxiv.org/abs/1004.0476}{{\ttfamily 1004.0476}}].

\bibitem{Bern:2019prr}
Z.~Bern, J.J.~Carrasco, M.~Chiodaroli, H.~Johansson and R.~Roiban, \emph{{The
  Duality Between Color and Kinematics and its Applications}},
  \href{https://arxiv.org/abs/1909.01358}{{\ttfamily 1909.01358}}.

\bibitem{Johansson:2015oia}
H.~Johansson and A.~Ochirov, \emph{{Color-Kinematics Duality for QCD
  Amplitudes}}, \href{https://doi.org/10.1007/JHEP01(2016)170}{\emph{JHEP}
  {\bfseries 01} (2016) 170}
  [\href{https://arxiv.org/abs/1507.00332}{{\ttfamily 1507.00332}}].

\bibitem{Ochirov:2018uyq}
A.~Ochirov, \emph{{Helicity amplitudes for QCD with massive quarks}},
  \href{https://doi.org/10.1007/JHEP04(2018)089}{\emph{JHEP} {\bfseries 04}
  (2018) 089} [\href{https://arxiv.org/abs/1802.06730}{{\ttfamily
  1802.06730}}].

\bibitem{Johansson:2019dnu}
H.~Johansson and A.~Ochirov, \emph{{Double copy for massive quantum particles
  with spin}}, \href{https://doi.org/10.1007/JHEP09(2019)040}{\emph{JHEP}
  {\bfseries 09} (2019) 040}
  [\href{https://arxiv.org/abs/1906.12292}{{\ttfamily 1906.12292}}].

\bibitem{Chung:2019duq}
M.-Z.~Chung, Y.-T.~Huang and J.-W.~Kim, \emph{{Classical potential for general
  spinning bodies}}, \href{https://doi.org/10.1007/JHEP09(2020)074}{\emph{JHEP}
  {\bfseries 09} (2020) 074}
  [\href{https://arxiv.org/abs/1908.08463}{{\ttfamily 1908.08463}}].

\bibitem{Bautista:2019evw}
Y.F.~Bautista and A.~Guevara, \emph{{On the double copy for spinning matter}},
  \href{https://doi.org/10.1007/JHEP11(2021)184}{\emph{JHEP} {\bfseries 11}
  (2021) 184} [\href{https://arxiv.org/abs/1908.11349}{{\ttfamily
  1908.11349}}].

\bibitem{Arkani-Hamed:2019ymq}
N.~Arkani-Hamed, Y.-t.~Huang and D.~O'Connell, \emph{{Kerr black holes as
  elementary particles}},
  \href{https://doi.org/10.1007/JHEP01(2020)046}{\emph{JHEP} {\bfseries 01}
  (2020) 046} [\href{https://arxiv.org/abs/1906.10100}{{\ttfamily
  1906.10100}}].

\bibitem{Edison:2020ehu}
A.~Edison and F.~Teng, \emph{{Efficient Calculation of Crossing Symmetric BCJ
  Tree Numerators}}, \href{https://doi.org/10.1007/JHEP12(2020)138}{\emph{JHEP}
  {\bfseries 12} (2020) 138}
  [\href{https://arxiv.org/abs/2005.03638}{{\ttfamily 2005.03638}}].

\bibitem{Johnson:2020pny}
L.A.~Johnson, C.R.T.~Jones and S.~Paranjape, \emph{{Constraints on a Massive
  Double-Copy and Applications to Massive Gravity}},
  \href{https://doi.org/10.1007/JHEP02(2021)148}{\emph{JHEP} {\bfseries 02}
  (2021) 148} [\href{https://arxiv.org/abs/2004.12948}{{\ttfamily
  2004.12948}}].

\bibitem{Momeni:2020hmc}
A.~Momeni, J.~Rumbutis and A.J.~Tolley, \emph{{Kaluza-Klein from
  colour-kinematics duality for massive fields}},
  \href{https://doi.org/10.1007/JHEP08(2021)081}{\emph{JHEP} {\bfseries 08}
  (2021) 081} [\href{https://arxiv.org/abs/2012.09711}{{\ttfamily
  2012.09711}}].

\bibitem{Momeni:2020vvr}
A.~Momeni, J.~Rumbutis and A.J.~Tolley, \emph{{Massive Gravity from Double
  Copy}}, \href{https://doi.org/10.1007/JHEP12(2020)030}{\emph{JHEP} {\bfseries
  12} (2020) 030} [\href{https://arxiv.org/abs/2004.07853}{{\ttfamily
  2004.07853}}].

\bibitem{Haddad:2020tvs}
K.~Haddad and A.~Helset, \emph{{The double copy for heavy particles}},
  \href{https://doi.org/10.1103/PhysRevLett.125.181603}{\emph{Phys. Rev. Lett.}
  {\bfseries 125} (2020) 181603}
  [\href{https://arxiv.org/abs/2005.13897}{{\ttfamily 2005.13897}}].

\bibitem{Emond:2020lwi}
W.T.~Emond, Y.-T.~Huang, U.~Kol, N.~Moynihan and D.~O'Connell,
  \emph{{Amplitudes from Coulomb to Kerr-Taub-NUT}},
  \href{https://doi.org/10.1007/JHEP05(2022)055}{\emph{JHEP} {\bfseries 05}
  (2022) 055} [\href{https://arxiv.org/abs/2010.07861}{{\ttfamily
  2010.07861}}].

\bibitem{Bjerrum-Bohr:2020syg}
N.E.J.~Bjerrum-Bohr, T.V.~Brown and H.~Gomez, \emph{{Scattering of Gravitons
  and Spinning Massive States from Compact Numerators}},
  \href{https://doi.org/10.1007/JHEP04(2021)234}{\emph{JHEP} {\bfseries 04}
  (2021) 234} [\href{https://arxiv.org/abs/2011.10556}{{\ttfamily
  2011.10556}}].

\bibitem{Brandhuber:2021kpo}
A.~Brandhuber, G.~Chen, G.~Travaglini and C.~Wen, \emph{{A new gauge-invariant
  double copy for heavy-mass effective theory}},
  \href{https://doi.org/10.1007/JHEP07(2021)047}{\emph{JHEP} {\bfseries 07}
  (2021) 047} [\href{https://arxiv.org/abs/2104.11206}{{\ttfamily
  2104.11206}}].

\bibitem{Monteiro:2014cda}
R.~Monteiro, D.~O'Connell and C.D.~White, \emph{{Black holes and the double
  copy}}, \href{https://doi.org/10.1007/JHEP12(2014)056}{\emph{JHEP} {\bfseries
  12} (2014) 056} [\href{https://arxiv.org/abs/1410.0239}{{\ttfamily
  1410.0239}}].

\bibitem{Huang:2019cja}
Y.-T.~Huang, U.~Kol and D.~O'Connell, \emph{{Double copy of electric-magnetic
  duality}}, \href{https://doi.org/10.1103/PhysRevD.102.046005}{\emph{Phys.
  Rev. D} {\bfseries 102} (2020) 046005}
  [\href{https://arxiv.org/abs/1911.06318}{{\ttfamily 1911.06318}}].

\bibitem{Chung:2019yfs}
M.-Z.~Chung, Y.-T.~Huang and J.-W.~Kim, \emph{{Kerr-Newman stress-tensor from
  minimal coupling}},
  \href{https://doi.org/10.1007/JHEP12(2020)103}{\emph{JHEP} {\bfseries 12}
  (2020) 103} [\href{https://arxiv.org/abs/1911.12775}{{\ttfamily
  1911.12775}}].

\bibitem{Guevara:2020xjx}
A.~Guevara, B.~Maybee, A.~Ochirov, D.~O'Connell and J.~Vines, \emph{{A
  worldsheet for Kerr}},
  \href{https://doi.org/10.1007/JHEP03(2021)201}{\emph{JHEP} {\bfseries 03}
  (2021) 201} [\href{https://arxiv.org/abs/2012.11570}{{\ttfamily
  2012.11570}}].

\bibitem{Maybee:2019jus}
B.~Maybee, D.~O'Connell and J.~Vines, \emph{{Observables and amplitudes for
  spinning particles and black holes}},
  \href{https://doi.org/10.1007/JHEP12(2019)156}{\emph{JHEP} {\bfseries 12}
  (2019) 156} [\href{https://arxiv.org/abs/1906.09260}{{\ttfamily
  1906.09260}}].

\bibitem{Aoude:2020onz}
R.~Aoude, K.~Haddad and A.~Helset, \emph{{On-shell heavy particle effective
  theories}}, \href{https://doi.org/10.1007/JHEP05(2020)051}{\emph{JHEP}
  {\bfseries 05} (2020) 051}
  [\href{https://arxiv.org/abs/2001.09164}{{\ttfamily 2001.09164}}].

\bibitem{Aoude:2021oqj}
R.~Aoude and A.~Ochirov, \emph{{Classical observables from coherent-spin
  amplitudes}}, \href{https://doi.org/10.1007/JHEP10(2021)008}{\emph{JHEP}
  {\bfseries 10} (2021) 008}
  [\href{https://arxiv.org/abs/2108.01649}{{\ttfamily 2108.01649}}].

\bibitem{Holstein:2008sw}
B.R.~Holstein and A.~Ross, \emph{{Spin Effects in Long Range Electromagnetic
  Scattering}},  \href{https://arxiv.org/abs/0802.0715}{{\ttfamily 0802.0715}}.

\bibitem{Holstein:2008sx}
B.R.~Holstein and A.~Ross, \emph{{Spin Effects in Long Range Gravitational
  Scattering}},  \href{https://arxiv.org/abs/0802.0716}{{\ttfamily 0802.0716}}.

\bibitem{Aoude:2020mlg}
R.~Aoude, M.-Z.~Chung, Y.-t.~Huang, C.S.~Machado and M.-K.~Tam, \emph{{Silence
  of Binary Kerr Black Holes}},
  \href{https://doi.org/10.1103/PhysRevLett.125.181602}{\emph{Phys. Rev. Lett.}
  {\bfseries 125} (2020) 181602}
  [\href{https://arxiv.org/abs/2007.09486}{{\ttfamily 2007.09486}}].

\bibitem{Chen:2021huj}
B.-T.~Chen, M.-Z.~Chung, Y.-t.~Huang and M.K.~Tam, \emph{{Minimal spin
  deflection of Kerr-Newman and supersymmetric black hole}},
  \href{https://doi.org/10.1007/JHEP10(2021)011}{\emph{JHEP} {\bfseries 10}
  (2021) 011} [\href{https://arxiv.org/abs/2106.12518}{{\ttfamily
  2106.12518}}].

\bibitem{Haddad:2021znf}
K.~Haddad, \emph{{Exponentiation of the leading eikonal phase with spin}},
  \href{https://doi.org/10.1103/PhysRevD.105.026004}{\emph{Phys. Rev. D}
  {\bfseries 105} (2022) 026004}
  [\href{https://arxiv.org/abs/2109.04427}{{\ttfamily 2109.04427}}].

\bibitem{Bern:2020buy}
Z.~Bern, A.~Luna, R.~Roiban, C.-H.~Shen and M.~Zeng, \emph{{Spinning black hole
  binary dynamics, scattering amplitudes, and effective field theory}},
  \href{https://doi.org/10.1103/PhysRevD.104.065014}{\emph{Phys. Rev. D}
  {\bfseries 104} (2021) 065014}
  [\href{https://arxiv.org/abs/2005.03071}{{\ttfamily 2005.03071}}].

\bibitem{Liu_2021}
Z.~Liu, R.A.~Porto and Z.~Yang, \emph{Spin effects in the effective field
  theory approach to post-minkowskian conservative dynamics},
  \href{https://doi.org/10.1007/jhep06(2021)012}{\emph{Journal of High Energy
  Physics} {\bfseries 2021} (2021) }.

\bibitem{Jakobsen:2021lvp}
G.U.~Jakobsen, G.~Mogull, J.~Plefka and J.~Steinhoff, \emph{{Gravitational
  Bremsstrahlung and Hidden Supersymmetry of Spinning Bodies}},
  \href{https://doi.org/10.1103/PhysRevLett.128.011101}{\emph{Phys. Rev. Lett.}
  {\bfseries 128} (2022) 011101}
  [\href{https://arxiv.org/abs/2106.10256}{{\ttfamily 2106.10256}}].

\bibitem{Jakobsen:2022fcj}
G.U.~Jakobsen and G.~Mogull, \emph{{Conservative and Radiative Dynamics of
  Spinning Bodies at Third Post-Minkowskian Order Using Worldline Quantum Field
  Theory}}, \href{https://doi.org/10.1103/PhysRevLett.128.141102}{\emph{Phys.
  Rev. Lett.} {\bfseries 128} (2022) 141102}
  [\href{https://arxiv.org/abs/2201.07778}{{\ttfamily 2201.07778}}].

\bibitem{Riva:2022fru}
M.M.~Riva, F.~Vernizzi and L.K.~Wong, \emph{{Gravitational Bremsstrahlung from
  Spinning Binaries in the Post-Minkowskian Expansion}},
  \href{https://arxiv.org/abs/2205.15295}{{\ttfamily 2205.15295}}.

\bibitem{Siemonsen:2019dsu}
N.~Siemonsen and J.~Vines, \emph{{Test black holes, scattering amplitudes and
  perturbations of Kerr spacetime}},
  \href{https://doi.org/10.1103/PhysRevD.101.064066}{\emph{Phys. Rev. D}
  {\bfseries 101} (2020) 064066}
  [\href{https://arxiv.org/abs/1909.07361}{{\ttfamily 1909.07361}}].

\bibitem{Damgaard:2019lfh}
P.H.~Damgaard, K.~Haddad and A.~Helset, \emph{{Heavy Black Hole Effective
  Theory}}, \href{https://doi.org/10.1007/JHEP11(2019)070}{\emph{JHEP}
  {\bfseries 11} (2019) 070}
  [\href{https://arxiv.org/abs/1908.10308}{{\ttfamily 1908.10308}}].

\bibitem{Kosmopoulos:2021zoq}
D.~Kosmopoulos and A.~Luna, \emph{{Quadratic-in-spin Hamiltonian at $
  \mathcal{O} $(G$^{2}$) from scattering amplitudes}},
  \href{https://doi.org/10.1007/JHEP07(2021)037}{\emph{JHEP} {\bfseries 07}
  (2021) 037} [\href{https://arxiv.org/abs/2102.10137}{{\ttfamily
  2102.10137}}].

\bibitem{Chen:2021qkk}
W.-M.~Chen, M.-Z.~Chung, Y.-t.~Huang and J.-W.~Kim, \emph{{The 2PM Hamiltonian
  for binary Kerr to quartic in spin}},
  \href{https://arxiv.org/abs/2111.13639}{{\ttfamily 2111.13639}}.

\bibitem{Menezes:2022tcs}
G.~Menezes and M.~Sergola, \emph{{NLO deflections for spinning particles and
  Kerr black holes}},  \href{https://arxiv.org/abs/2205.11701}{{\ttfamily
  2205.11701}}.

\bibitem{FebresCordero:2022jts}
F.~Febres~Cordero, M.~Kraus, G.~Lin, M.S.~Ruf and M.~Zeng, \emph{{Conservative
  Binary Dynamics with a Spinning Black Hole at $\mathcal{O}(G^3)$ from
  Scattering Amplitudes}},  \href{https://arxiv.org/abs/2205.07357}{{\ttfamily
  2205.07357}}.

\bibitem{Aoude:2020ygw}
R.~Aoude, K.~Haddad and A.~Helset, \emph{{Tidal effects for spinning
  particles}}, \href{https://doi.org/10.1007/JHEP03(2021)097}{\emph{JHEP}
  {\bfseries 03} (2021) 097}
  [\href{https://arxiv.org/abs/2012.05256}{{\ttfamily 2012.05256}}].

\bibitem{Bern:2020uwk}
Z.~Bern, J.~Parra-Martinez, R.~Roiban, E.~Sawyer and C.-H.~Shen, \emph{{Leading
  Nonlinear Tidal Effects and Scattering Amplitudes}},
  \href{https://doi.org/10.1007/JHEP05(2021)188}{\emph{JHEP} {\bfseries 05}
  (2021) 188} [\href{https://arxiv.org/abs/2010.08559}{{\ttfamily
  2010.08559}}].

\bibitem{Mogull:2020sak}
G.~Mogull, J.~Plefka and J.~Steinhoff, \emph{{Classical black hole scattering
  from a worldline quantum field theory}},
  \href{https://doi.org/10.1007/JHEP02(2021)048}{\emph{JHEP} {\bfseries 02}
  (2021) 048} [\href{https://arxiv.org/abs/2010.02865}{{\ttfamily
  2010.02865}}].

\bibitem{AccettulliHuber:2020dal}
M.~Accettulli~Huber, A.~Brandhuber, S.~De~Angelis and G.~Travaglini,
  \emph{{From amplitudes to gravitational radiation with cubic interactions and
  tidal effects}},
  \href{https://doi.org/10.1103/PhysRevD.103.045015}{\emph{Phys. Rev. D}
  {\bfseries 103} (2021) 045015}
  [\href{https://arxiv.org/abs/2012.06548}{{\ttfamily 2012.06548}}].

\bibitem{Cristofoli:2021vyo}
A.~Cristofoli, R.~Gonzo, D.A.~Kosower and D.~O'Connell, \emph{{Waveforms from
  Amplitudes}},  \href{https://arxiv.org/abs/2107.10193}{{\ttfamily
  2107.10193}}.

\bibitem{Herrmann:2021lqe}
E.~Herrmann, J.~Parra-Martinez, M.S.~Ruf and M.~Zeng, \emph{{Gravitational
  Bremsstrahlung from Reverse Unitarity}},
  \href{https://doi.org/10.1103/PhysRevLett.126.201602}{\emph{Phys. Rev. Lett.}
  {\bfseries 126} (2021) 201602}
  [\href{https://arxiv.org/abs/2101.07255}{{\ttfamily 2101.07255}}].

\bibitem{Mougiakakos:2021ckm}
S.~Mougiakakos, M.M.~Riva and F.~Vernizzi, \emph{{Gravitational Bremsstrahlung
  in the post-Minkowskian effective field theory}},
  \href{https://doi.org/10.1103/PhysRevD.104.024041}{\emph{Phys. Rev. D}
  {\bfseries 104} (2021) 024041}
  [\href{https://arxiv.org/abs/2102.08339}{{\ttfamily 2102.08339}}].

\bibitem{Bautista:2021inx}
Y.F.~Bautista and N.~Siemonsen, \emph{{Post-Newtonian waveforms from spinning
  scattering amplitudes}},
  \href{https://doi.org/10.1007/JHEP01(2022)006}{\emph{JHEP} {\bfseries 01}
  (2022) 006} [\href{https://arxiv.org/abs/2110.12537}{{\ttfamily
  2110.12537}}].

\bibitem{Alessio:2022kwv}
F.~Alessio and P.~Di~Vecchia, \emph{{Radiation reaction for spinning black-hole
  scattering}},  \href{https://arxiv.org/abs/2203.13272}{{\ttfamily
  2203.13272}}.

\bibitem{Bekaert:2022poo}
X.~Bekaert, N.~Boulanger, A.~Campoleoni, M.~Chiodaroli, D.~Francia,
  M.~Grigoriev et~al., \emph{{Snowmass White Paper: Higher Spin Gravity and
  Higher Spin symmetry}},  \href{https://arxiv.org/abs/2205.01567}{{\ttfamily
  2205.01567}}.

\bibitem{Britto:2004ap}
R.~Britto, F.~Cachazo and B.~Feng, \emph{{New recursion relations for tree
  amplitudes of gluons}},
  \href{https://doi.org/10.1016/j.nuclphysb.2005.02.030}{\emph{Nucl. Phys. B}
  {\bfseries 715} (2005) 499}
  [\href{https://arxiv.org/abs/hep-th/0412308}{{\ttfamily hep-th/0412308}}].

\bibitem{Britto:2005fq}
R.~Britto, F.~Cachazo, B.~Feng and E.~Witten, \emph{{Direct proof of tree-level
  recursion relation in Yang-Mills theory}},
  \href{https://doi.org/10.1103/PhysRevLett.94.181602}{\emph{Phys. Rev. Lett.}
  {\bfseries 94} (2005) 181602}
  [\href{https://arxiv.org/abs/hep-th/0501052}{{\ttfamily hep-th/0501052}}].

\bibitem{Aoude:2022trd}
R.~Aoude, K.~Haddad and A.~Helset, \emph{{Searching for Kerr in the 2PM
  amplitude}},  \href{https://arxiv.org/abs/2203.06197}{{\ttfamily
  2203.06197}}.

\bibitem{Bern:2022kto}
Z.~Bern, D.~Kosmopoulos, A.~Luna, R.~Roiban and F.~Teng, \emph{{Binary Dynamics
  Through the Fifth Power of Spin at $\mathcal{O}(G^2)$}},
  \href{https://arxiv.org/abs/2203.06202}{{\ttfamily 2203.06202}}.

\bibitem{Aoude:2022thd}
R.~Aoude, K.~Haddad and A.~Helset, \emph{{Classical gravitational
  spinning-spinless scattering at $\mathcal{O}(G^{2} S^{\infty})$}},
  \href{https://arxiv.org/abs/2205.02809}{{\ttfamily 2205.02809}}.

\bibitem{Chen:2022yxw}
W.-M.~Chen, M.-Z.~Chung, Y.-t.~Huang and J.-W.~Kim, \emph{{Lense-Thirring
  effects from on-shell amplitudes}},
  \href{https://arxiv.org/abs/2205.07305}{{\ttfamily 2205.07305}}.

\bibitem{Chiodaroli:2021eug}
M.~Chiodaroli, H.~Johansson and P.~Pichini, \emph{{Compton black-hole
  scattering for s \ensuremath{\leq} 5/2}},
  \href{https://doi.org/10.1007/JHEP02(2022)156}{\emph{JHEP} {\bfseries 02}
  (2022) 156} [\href{https://arxiv.org/abs/2107.14779}{{\ttfamily
  2107.14779}}].

\bibitem{Falkowski:2020aso}
A.~Falkowski and C.S.~Machado, \emph{{Soft Matters, or the Recursions with
  Massive Spinors}}, \href{https://doi.org/10.1007/JHEP05(2021)238}{\emph{JHEP}
  {\bfseries 05} (2021) 238}
  [\href{https://arxiv.org/abs/2005.08981}{{\ttfamily 2005.08981}}].

\bibitem{Bautista:2021wfy}
Y.F.~Bautista, A.~Guevara, C.~Kavanagh and J.~Vines, \emph{{From Scattering in
  Black Hole Backgrounds to Higher-Spin Amplitudes: Part I}},
  \href{https://arxiv.org/abs/2107.10179}{{\ttfamily 2107.10179}}.

\bibitem{KOH1987201}
I.~Koh, W.~Troost and A.~{Van Proeyen}, \emph{Covariant higher spin vertex
  operators in the ramond sector},
  \href{https://doi.org/https://doi.org/10.1016/0550-3213(87)90642-0}{\emph{Nuclear
  Physics B} {\bfseries 292} (1987) 201}.

\bibitem{Berkovits:2002qx}
N.~Berkovits and O.~Chandia, \emph{{Massive superstring vertex operator in D =
  10 superspace}},
  \href{https://doi.org/10.1088/1126-6708/2002/08/040}{\emph{JHEP} {\bfseries
  08} (2002) 040} [\href{https://arxiv.org/abs/hep-th/0204121}{{\ttfamily
  hep-th/0204121}}].

\bibitem{Hanany:2010da}
A.~Hanany, D.~Forcella and J.~Troost, \emph{{The Covariant perturbative string
  spectrum}},
  \href{https://doi.org/10.1016/j.nuclphysb.2011.01.002}{\emph{Nucl. Phys. B}
  {\bfseries 846} (2011) 212}
  [\href{https://arxiv.org/abs/1007.2622}{{\ttfamily 1007.2622}}].

\bibitem{Bianchi:2010es}
M.~Bianchi, L.~Lopez and R.~Richter, \emph{{On stable higher spin states in
  Heterotic String Theories}},
  \href{https://doi.org/10.1007/JHEP03(2011)051}{\emph{JHEP} {\bfseries 03}
  (2011) 051} [\href{https://arxiv.org/abs/1010.1177}{{\ttfamily 1010.1177}}].

\bibitem{Lust:2012zv}
D.~L\"ust, N.~Mekareeya, O.~Schlotterer and A.~Thomson, \emph{{Refined
  Partition Functions for Open Superstrings with 4, 8 and 16 Supercharges}},
  \href{https://doi.org/10.1016/j.nuclphysb.2013.08.003}{\emph{Nucl. Phys. B}
  {\bfseries 876} (2013) 55} [\href{https://arxiv.org/abs/1211.1018}{{\ttfamily
  1211.1018}}].

\bibitem{Lee:2012fu}
J.-C.~Lee and Y.~Mitsuka, \emph{{Recurrence relations of Kummer functions and
  Regge string scattering amplitudes}},
  \href{https://doi.org/10.1007/JHEP04(2013)082}{\emph{JHEP} {\bfseries 04}
  (2013) 082} [\href{https://arxiv.org/abs/1212.6915}{{\ttfamily 1212.6915}}].

\bibitem{Feng:2012bb}
W.-Z.~Feng, D.~Lust and O.~Schlotterer, \emph{{Massive Supermultiplets in
  Four-Dimensional Superstring Theory}},
  \href{https://doi.org/10.1016/j.nuclphysb.2012.03.010}{\emph{Nucl. Phys. B}
  {\bfseries 861} (2012) 175}
  [\href{https://arxiv.org/abs/1202.4466}{{\ttfamily 1202.4466}}].

\bibitem{Fu:2013xba}
C.-H.~Fu, J.-C.~Lee, C.-I.~Tan and Y.~Yang, \emph{{Recurrence relations of
  higher spin BPST vertex operators for open strings}},
  \href{https://doi.org/10.1103/PhysRevD.88.046004}{\emph{Phys. Rev. D}
  {\bfseries 88} (2013) 046004}
  [\href{https://arxiv.org/abs/1304.6948}{{\ttfamily 1304.6948}}].

\bibitem{PhysRevD.83.046005}
D.~Polyakov, \emph{Higher spins and open strings: Quartic interactions},
  \href{https://doi.org/10.1103/PhysRevD.83.046005}{\emph{Phys. Rev. D}
  {\bfseries 83} (2011) 046005}.

\bibitem{Feng:2010yx}
W.-Z.~Feng, D.~Lust, O.~Schlotterer, S.~Stieberger and T.R.~Taylor,
  \emph{{Direct Production of Lightest Regge Resonances}},
  \href{https://doi.org/10.1016/j.nuclphysb.2010.10.013}{\emph{Nucl. Phys. B}
  {\bfseries 843} (2011) 570}
  [\href{https://arxiv.org/abs/1007.5254}{{\ttfamily 1007.5254}}].

\bibitem{Schlotterer:2011zz}
O.~Schlotterer, \emph{{SUSY multiplets at first mass level in D=4 superstring
  compactifications}},
  \href{https://doi.org/10.1016/j.nuclphysbps.2011.05.009}{\emph{Nucl. Phys. B
  Proc. Suppl.} {\bfseries 216} (2011) 265}.

\bibitem{Feng:2011qc}
W.-Z.~Feng and T.R.~Taylor, \emph{{Higher Level String Resonances in Four
  Dimensions}},
  \href{https://doi.org/10.1016/j.nuclphysb.2011.11.004}{\emph{Nucl. Phys. B}
  {\bfseries 856} (2012) 247}
  [\href{https://arxiv.org/abs/1110.1087}{{\ttfamily 1110.1087}}].

\bibitem{Boels:2012if}
R.H.~Boels, \emph{{Three particle superstring amplitudes with massive legs}},
  \href{https://doi.org/10.1007/JHEP06(2012)026}{\emph{JHEP} {\bfseries 06}
  (2012) 026} [\href{https://arxiv.org/abs/1201.2655}{{\ttfamily 1201.2655}}].

\bibitem{Tsulaia:2012rb}
M.~Tsulaia, \emph{{On Tensorial Spaces and BCFW Recursion Relations for Higher
  Spin Fields}}, \href{https://doi.org/10.1142/S0217751X12300116}{\emph{Int. J.
  Mod. Phys. A} {\bfseries 27} (2012) 1230011}
  [\href{https://arxiv.org/abs/1202.6309}{{\ttfamily 1202.6309}}].

\bibitem{Feng:2012qia}
W.-Z.~Feng, \emph{{Physics of massive superstrings}}, Ph.D. thesis,
  Northeastern U., 2012.

\bibitem{Boels:2014dka}
R.H.~Boels and T.~Hansen, \emph{{String theory in target space}},
  \href{https://doi.org/10.1007/JHEP06(2014)054}{\emph{JHEP} {\bfseries 06}
  (2014) 054} [\href{https://arxiv.org/abs/1402.6356}{{\ttfamily 1402.6356}}].

\bibitem{Minahan:2014usa}
J.A.~Minahan and R.~Pereira, \emph{{Three-point correlators from string
  amplitudes: Mixing and Regge spins}},
  \href{https://doi.org/10.1007/JHEP04(2015)134}{\emph{JHEP} {\bfseries 04}
  (2015) 134} [\href{https://arxiv.org/abs/1410.4746}{{\ttfamily 1410.4746}}].

\bibitem{Hansen:2015pqa}
T.~Hansen, \emph{{Dissecting CFT Correlators and String Amplitudes: Conformal
  Blocks and On-Shell Recursion for General Tensor Fields}}, Ph.D. thesis,
  Hamburg U., 2015.
\newblock 10.3204/DESY-THESIS-2015-026.

\bibitem{Chakrabarti:2018bah}
S.~Chakrabarti, S.P.~Kashyap and M.~Verma, \emph{{Amplitudes Involving Massive
  States Using Pure Spinor Formalism}},
  \href{https://doi.org/10.1007/JHEP12(2018)071}{\emph{JHEP} {\bfseries 12}
  (2018) 071} [\href{https://arxiv.org/abs/1808.08735}{{\ttfamily
  1808.08735}}].

\bibitem{Lee:2019ufv}
T.~Lee and H.~Park, \emph{{Graviton and Massive Symmetric Rank-Two Tensor in
  String Theory}},
  \href{https://doi.org/10.5506/APhysPolBSupp.13.303}{\emph{Acta Phys. Polon.
  Supp.} {\bfseries 13} (2020) 303}
  [\href{https://arxiv.org/abs/1909.08516}{{\ttfamily 1909.08516}}].

\bibitem{Gross:2021gsj}
D.J.~Gross and V.~Rosenhaus, \emph{{Chaotic scattering of highly excited
  strings}}, \href{https://doi.org/10.1007/JHEP05(2021)048}{\emph{JHEP}
  {\bfseries 05} (2021) 048}
  [\href{https://arxiv.org/abs/2103.15301}{{\ttfamily 2103.15301}}].

\bibitem{Jusinskas:2021bdj}
R.L.~Jusinskas, \emph{{Asymmetrically twisted strings}},
  \href{https://doi.org/10.1016/j.physletb.2022.137090}{\emph{Phys. Lett. B}
  {\bfseries 829} (2022) 137090}
  [\href{https://arxiv.org/abs/2108.13426}{{\ttfamily 2108.13426}}].

\bibitem{Guillen:2021mwp}
M.~Guillen, H.~Johansson, R.L.~Jusinskas and O.~Schlotterer, \emph{{Scattering
  Massive String Resonances through Field-Theory Methods}},
  \href{https://doi.org/10.1103/PhysRevLett.127.051601}{\emph{Phys. Rev. Lett.}
  {\bfseries 127} (2021) 051601}
  [\href{https://arxiv.org/abs/2104.03314}{{\ttfamily 2104.03314}}].

\bibitem{Giannakis:1998wi}
I.~Giannakis, J.T.~Liu and M.~Porrati, \emph{{Massive higher spin states in
  string theory and the principle of equivalence}},
  \href{https://doi.org/10.1103/PhysRevD.59.104013}{\emph{Phys. Rev. D}
  {\bfseries 59} (1999) 104013}
  [\href{https://arxiv.org/abs/hep-th/9809142}{{\ttfamily hep-th/9809142}}].

\bibitem{Buchbinder:1999be}
I.L.~Buchbinder, V.A.~Krykhtin and V.D.~Pershin, \emph{{On consistent equations
  for massive spin two field coupled to gravity in string theory}},
  \href{https://doi.org/10.1016/S0370-2693(99)01143-0}{\emph{Phys. Lett. B}
  {\bfseries 466} (1999) 216}
  [\href{https://arxiv.org/abs/hep-th/9908028}{{\ttfamily hep-th/9908028}}].

\bibitem{Buchbinder:1999ar}
I.L.~Buchbinder, D.M.~Gitman, V.A.~Krykhtin and V.D.~Pershin, \emph{{Equations
  of motion for massive spin-2 field coupled to gravity}},
  \href{https://doi.org/10.1016/S0550-3213(00)00389-8}{\emph{Nucl. Phys. B}
  {\bfseries 584} (2000) 615}
  [\href{https://arxiv.org/abs/hep-th/9910188}{{\ttfamily hep-th/9910188}}].

\bibitem{Buchbinder:2000ta}
I.L.~Buchbinder and V.D.~Pershin, \emph{{Gravitational interaction of higher
  spin massive fields and string theory}},  in \emph{{Conference on Geometrical
  Aspects of Quantum Fields}}, pp.~11--30, 4, 2000,
  \href{https://doi.org/10.1142/9789812810366_0002}{DOI}
  [\href{https://arxiv.org/abs/hep-th/0009026}{{\ttfamily hep-th/0009026}}].

\bibitem{PhysRev.186.1337}
G.~Velo and D.~Zwanziger, \emph{Propagation and quantization of
  rarita-schwinger waves in an external electromagnetic potential},
  \href{https://doi.org/10.1103/PhysRev.186.1337}{\emph{Phys. Rev.} {\bfseries
  186} (1969) 1337}.

\bibitem{PhysRev.188.2218}
G.~Velo and D.~Zwanzinger, \emph{Noncausality and other defects of interaction
  lagrangians for particles with spin one and higher},
  \href{https://doi.org/10.1103/PhysRev.188.2218}{\emph{Phys. Rev.} {\bfseries
  188} (1969) 2218}.

\bibitem{Taronna:2010qq}
M.~Taronna, \emph{{Higher Spins and String Interactions}},  other thesis,
  Scuola Normale Superiore - PISA, 5, 2010,
  [\href{https://arxiv.org/abs/1005.3061}{{\ttfamily 1005.3061}}].

\bibitem{Joung:2012rv}
E.~Joung, L.~Lopez and M.~Taronna, \emph{{On the cubic interactions of massive
  and partially-massless higher spins in (A)dS}},
  \href{https://doi.org/10.1007/JHEP07(2012)041}{\emph{JHEP} {\bfseries 07}
  (2012) 041} [\href{https://arxiv.org/abs/1203.6578}{{\ttfamily 1203.6578}}].

\bibitem{Taronna:2012gb}
M.~Taronna, \emph{{Higher-Spin Interactions: three-point functions and
  beyond}}, Ph.D. thesis, Pisa, Scuola Normale Superiore, 2012.
\newblock \href{https://arxiv.org/abs/1209.5755}{{\ttfamily 1209.5755}}.

\bibitem{Rahman:2015pzl}
R.~Rahman and M.~Taronna, \emph{{From Higher Spins to Strings: A Primer}},
  \href{https://arxiv.org/abs/1512.07932}{{\ttfamily 1512.07932}}.

\bibitem{Marotta:2021oiw}
R.~Marotta, M.~Taronna and M.~Verma, \emph{{Revisiting higher-spin gyromagnetic
  couplings}}, \href{https://doi.org/10.1007/JHEP06(2021)168}{\emph{JHEP}
  {\bfseries 06} (2021) 168}
  [\href{https://arxiv.org/abs/2102.13180}{{\ttfamily 2102.13180}}].

\bibitem{Gross:1988ue}
D.J.~Gross, \emph{{High-Energy Symmetries of String Theory}},
  \href{https://doi.org/10.1103/PhysRevLett.60.1229}{\emph{Phys. Rev. Lett.}
  {\bfseries 60} (1988) 1229}.

\bibitem{Lee:1994fy}
J.-C.~Lee, \emph{{Spontaneously broken symmetry in string theory}},
  \href{https://doi.org/10.1016/0370-2693(94)91195-9}{\emph{Phys. Lett. B}
  {\bfseries 326} (1994) 79}
  [\href{https://arxiv.org/abs/hep-th/0503056}{{\ttfamily hep-th/0503056}}].

\bibitem{Chan:2003ee}
C.-T.~Chan and J.-C.~Lee, \emph{{Stringy symmetries and their high-energy
  limits}}, \href{https://doi.org/10.1016/j.physletb.2005.02.034}{\emph{Phys.
  Lett. B} {\bfseries 611} (2005) 193}
  [\href{https://arxiv.org/abs/hep-th/0312226}{{\ttfamily hep-th/0312226}}].

\bibitem{Chan:2004tb}
C.-T.~Chan, P.-M.~Ho and J.-C.~Lee, \emph{{Ward identities and high-energy
  scattering amplitudes in string theory}},
  \href{https://doi.org/10.1016/j.nuclphysb.2004.11.032}{\emph{Nucl. Phys. B}
  {\bfseries 708} (2005) 99}
  [\href{https://arxiv.org/abs/hep-th/0410194}{{\ttfamily hep-th/0410194}}].

\bibitem{Chan:2005ne}
C.-T.~Chan, P.-M.~Ho, J.-C.~Lee, S.~Teraguchi and Y.~Yang, \emph{{Solving all
  4-point correlation functions for bosonic open string theory in the high
  energy limit}},
  \href{https://doi.org/10.1016/j.nuclphysb.2005.07.018}{\emph{Nucl. Phys. B}
  {\bfseries 725} (2005) 352}
  [\href{https://arxiv.org/abs/hep-th/0504138}{{\ttfamily hep-th/0504138}}].

\bibitem{Chan:2005zp}
C.-T.~Chan, P.-M.~Ho, J.-C.~Lee, S.~Teraguchi and Y.~Yang, \emph{{High-energy
  zero-norm states and symmetries of string theory}},
  \href{https://doi.org/10.1103/PhysRevLett.96.171601}{\emph{Phys. Rev. Lett.}
  {\bfseries 96} (2006) 171601}
  [\href{https://arxiv.org/abs/hep-th/0505035}{{\ttfamily hep-th/0505035}}].

\bibitem{Chan:2006qf}
C.-T.~Chan, J.-C.~Lee and Y.~Yang, \emph{{Scatterings of Massive String States
  from D-brane and Their Linear Relations at High Energies}},
  \href{https://doi.org/10.1016/j.nuclphysb.2006.11.014}{\emph{Nucl. Phys. B}
  {\bfseries 764} (2007) 1}
  [\href{https://arxiv.org/abs/hep-th/0610062}{{\ttfamily hep-th/0610062}}].

\bibitem{Polyakov:2009pk}
D.~Polyakov, \emph{{Interactions of Massless Higher Spin Fields From String
  Theory}}, \href{https://doi.org/10.1103/PhysRevD.82.066005}{\emph{Phys. Rev.
  D} {\bfseries 82} (2010) 066005}
  [\href{https://arxiv.org/abs/0910.5338}{{\ttfamily 0910.5338}}].

\bibitem{Polyakov:2010qs}
D.~Polyakov, \emph{{Gravitational Couplings of Higher Spins from String
  Theory}}, \href{https://doi.org/10.1142/S0217751X1005041X}{\emph{Int. J. Mod.
  Phys. A} {\bfseries 25} (2010) 4623}
  [\href{https://arxiv.org/abs/1005.5512}{{\ttfamily 1005.5512}}].

\bibitem{Fotopoulos:2010ay}
A.~Fotopoulos and M.~Tsulaia, \emph{{On the Tensionless Limit of String theory,
  Off - Shell Higher Spin Interaction Vertices and BCFW Recursion Relations}},
  \href{https://doi.org/10.1007/JHEP11(2010)086}{\emph{JHEP} {\bfseries 11}
  (2010) 086} [\href{https://arxiv.org/abs/1009.0727}{{\ttfamily 1009.0727}}].

\bibitem{Bianchi:2011se}
M.~Bianchi and P.~Teresi, \emph{{Scattering higher spins off D-branes}},
  \href{https://doi.org/10.1007/JHEP01(2012)161}{\emph{JHEP} {\bfseries 01}
  (2012) 161} [\href{https://arxiv.org/abs/1108.1071}{{\ttfamily 1108.1071}}].

\bibitem{Lee:2019lvp}
J.-C.~Lee and Y.~Yang, \emph{{Overview of High Energy String Scattering
  Amplitudes and Symmetries of String Theory}},
  \href{https://doi.org/10.3390/sym11081045}{\emph{Symmetry} {\bfseries 11}
  (2019) 1045} [\href{https://arxiv.org/abs/1907.12810}{{\ttfamily
  1907.12810}}].

\bibitem{Sagnotti:2010at}
A.~Sagnotti and M.~Taronna, \emph{{String Lessons for Higher-Spin
  Interactions}},
  \href{https://doi.org/10.1016/j.nuclphysb.2010.08.019}{\emph{Nucl. Phys. B}
  {\bfseries 842} (2011) 299}
  [\href{https://arxiv.org/abs/1006.5242}{{\ttfamily 1006.5242}}].

\bibitem{Schlotterer:2010kk}
O.~Schlotterer, \emph{{Higher Spin Scattering in Superstring Theory}},
  \href{https://doi.org/10.1016/j.nuclphysb.2011.03.026}{\emph{Nucl. Phys. B}
  {\bfseries 849} (2011) 433}
  [\href{https://arxiv.org/abs/1011.1235}{{\ttfamily 1011.1235}}].

\bibitem{Ademollo:1974te}
M.~Ademollo, A.~D'Adda, R.~D'Auria, E.~Napolitano, S.~Sciuto, P.~Di~Vecchia
  et~al., \emph{{Theory of an interacting string and dual resonance model}},
  \href{https://doi.org/10.1007/BF02731188}{\emph{Nuovo Cim. A} {\bfseries 21}
  (1974) 77}.

\bibitem{Lust1989}
D.~Lust, \emph{The classical bosonic string},  in \emph{Lectures on String
  Theory}, (Berlin, Heidelberg), pp.~5--30, Springer Berlin Heidelberg (1989),
  \href{https://doi.org/10.1007/BFb0113509}{DOI}.

\bibitem{zwiebach_2004}
B.~Zwiebach, \emph{A First Course in String Theory}, Cambridge University Press
  (2004),
  \href{https://doi.org/10.1017/CBO9780511841682}{10.1017/CBO9780511841682}.

\bibitem{Bianchi:2019ywd}
M.~Bianchi and M.~Firrotta, \emph{{DDF operators, open string coherent states
  and their scattering amplitudes}},
  \href{https://doi.org/10.1016/j.nuclphysb.2020.114943}{\emph{Nucl. Phys. B}
  {\bfseries 952} (2020) 114943}
  [\href{https://arxiv.org/abs/1902.07016}{{\ttfamily 1902.07016}}].

\bibitem{green_schwarz_witten_2012}
M.B.~Green, J.H.~Schwarz and E.~Witten, \emph{Superstring Theory: 25th
  Anniversary Edition}, vol.~1 of \emph{Cambridge Monographs on Mathematical
  Physics}, Cambridge University Press (2012),
  \href{https://doi.org/10.1017/CBO9781139248563}{10.1017/CBO9781139248563}.

\bibitem{polchinski_1998}
J.~Polchinski, \emph{String Theory}, vol.~1 of \emph{Cambridge Monographs on
  Mathematical Physics}, Cambridge University Press (1998),
  \href{https://doi.org/10.1017/CBO9780511816079}{10.1017/CBO9780511816079}.

\bibitem{Tong:2009np}
D.~Tong, \emph{{String Theory}},
  \href{https://arxiv.org/abs/0908.0333}{{\ttfamily 0908.0333}}.

\end{thebibliography}\endgroup
\end{document}